\documentclass{iopjournal}

\usepackage{graphicx}
\usepackage{dcolumn}
\usepackage{xcolor}
\usepackage{amsfonts}
\usepackage{array}
\usepackage{booktabs}
\usepackage{tabularx}
\usepackage{url}
\usepackage{amsmath}
\usepackage{amssymb}
\usepackage{mathrsfs}
\usepackage{mathtools}
\usepackage{bm}
\usepackage{microtype}
\usepackage{ragged2e}
\AtBeginDocument{\justifying}

\usepackage{hyperref}

\newcommand{\dd}{\mathop{}\!\mathrm{d}}
\newcommand{\ket}[1]{\lvert #1 \rangle}

\newcommand{\blue}[1]{{\color{black}#1}}

\newcommand{\iu}{\mathrm{i}\mkern1mu}

\let\vec=\mathbf

\begin{document}

\articletype{Paper}

\title{Geometric Phases and Holonomy in Structured Optical Fields --- a tutorial}

\author{Kristina Frizyuk$^{1,*}$, Evgenii Menshikov$^2$ and Mauro Spera$^3$}

\affil{$^1$Institute of Theoretical Solid State Physics, Karlsruhe Institute of Technology, Karlsruhe, Germany}

\affil{$^2$Department of Information Engineering, University of Brescia, Brescia, Italy}

\affil{$^3$Dipartimento di Matematica e Fisica ``Niccolò Tartaglia'', Università Cattolica del Sacro Cuore, Brescia, Italy}

\affil{$^*$Author to whom any correspondence should be addressed.}

\email{kristina.friziuk@kit.edu}


\begin{abstract}
    {Geometric phases are widely used in modern optics, yet their meaning and underlying geometry depend on the actual physical settings, which can substantially differ from one another. This tutorial article introduces geometric phases in nanophotonic systems, focusing on the interaction of  structured light with nanostructures or metaatoms. 
    \blue{We compare the present setting with conventional geometric phases of structured-light optics and show that similar phase laws may correspond to genuinely different underlying geometries.}
    Our aim is to provide a pedagogical bridge between the mathematical language of geometric phases and experimentally relevant examples from nanophotonics.}
\end{abstract}
\tableofcontents
\section{Introduction}
Geometric phases are widely used in optics~\cite{Cisowski2022-Colloquium, Cohen2019-Geometricphasefrom} for different applications. 
The simplest one is transmission of a polarized plane wave through a waveplate~\cite{Pancharatnam1956-Generalizedtheoryof, Berry1987-TheAdiabaticPhasea}, corresponding 
to the rotation of Poincar\'e sphere around a horizontal axis.
\blue{Among the best-known examples we find the Pancharatnam--Berry metasurfaces, which use the orientation of anisotropic metaatoms to control the phase of the cross-polarized component~\cite{Ding2015-UltrathinPancharatna, Luo2017-TransmissiveUltrathi, Huang2012-DispersionlessPhase}. 
This principle has enabled dielectric metalenses~\cite{doi:10.1126/science.aaf6644} and reflective metasurface holograms~\cite{Wen2015-HelicityMultiplexed, Zheng2015-MetasurfaceHolograms}. Patterned liquid-crystal elements provide a related implementation~\cite{Marrucci2006May}, while metaatoms with higher-fold rotational symmetries can realize generalized Pancharatnam--Berry phases~\cite{Xie2021-GeneralizedPancharat, DENG2023106730}.}

\blue{Geometric phases can also be introduced for beams with angular momentum through the associated Poincar\'e-sphere representation~\cite{Galvez2003-GeometricPhaseAssoc, Padgett1999-Poincare-sphereequiv, vanEnk1993-Geometricphasetran}. Their extension to higher-order transverse modes is discussed, for example, in~\cite{Habraken2010-Universaldescription, Habraken2010-Geometricphasesinh}, see also the review~\cite{Cisowski2022-Colloquium}.} 
\blue{Geometric phases arise in distinct optical settings and need not share the same underlying geometry. Besides the conventional Pancharatnam--Berry phase of polarization evolution, they occur in polarization transport along curved optical trajectories~\cite{Tomita1986,Bliokh2008Geometrodynamics}, in polarization-dependent beam shifts and reflected structured fields~\cite{Zhen2023,Zhen2025}, and in geometrically twisted optical resonators~\cite{Wang2023Mobius}. Distinct geometrical structures also arise for fixed-helicity states in momentum space~\cite{Palmerduca2024Apr} and for higher-order spatial modes on Hermite--Laguerre modal spheres~\cite{Cao2024}.}

The following
question 
then arises:
why are these particular types of phases called ``geometric''?
What is then the underlying geometry? 
This question has been initially addressed in~\cite{Simon1983-HolonomytheQuantum}, which connects the phase, introduced by Berry with the holonomy of a connection on a line bundle. 
{Here, we discuss the geometric phases for vortex beams from this point of view, via an explicit description of the underlying topological and geometrical objects, \blue{comparing two different types of possible phases}.}
{The effect enables mode-selective phase control and discrimination of vortex beams based on their total angular momentum (TAM), using simple nanostructures, offering a practical alternative to spatial light modulators and multi-element polarization schemes.}

We first start with the description of the simplest well-known case of a waveplate and circularly polarized plane waves, and later generalize it by \blue{introducing an analogous geometric phase for optical beams.}
Then we provide a mathematical description of the well-known geometric phase for beams, showing the emergence of a different geometry.

{
\blue{The paper is organized as follows: in Section~\ref{sec:simplest} we revisit the well-known Pancharatnam geometric phase and describe its geometry in detail.
In Section~\ref{sec:mainres} we describe a direct Pancharatnam--Berry-phase analogy for vortex beams.
In Section~\ref{sec:HG_beams} we discuss the usual geometric phase of higher-order transverse modes and show that its underlying geometry is different.
In Section~\ref{sec:discussion}, we compare these geometrical descriptions and discuss open conceptual issues concerning geometric phases in linear and nonlinear optics.}

{The guiding point of this work is the observation that in order to assign a geometric or topological meaning to an optical phase, one must first specify the bundle structure over the relevant base space~\cite{Palmerduca2024Apr, Cayssol2021-Topologicalandgeome}. 
Only then \blue{can} notions such as connection, holonomy, curvature, or topological invariants be accommodated and acquire a precise meaning. 
Accordingly, we focus not only on the accumulated phase, but also on the underlying geometry.

This tutorial is intended both for readers working in structured light, nanophotonics, and optical angular momentum, and for mathematically oriented readers interested in concrete physical examples of geometric phases. 
Polarization optics offers perhaps one of the simplest experimentally accessible, yet mathematically nontrivial, settings in which the notions of state space, base space, connection, and holonomy can be observed explicitly. 
We use this example as a starting point for comparing related optical phase effects and clarifying the geometrical structures behind them.}

\blue{Recent reviews cover the historical development and experimental realizations of geometric phases~\cite{Rao2026}, as well as wavefront control and waveguiding~\cite{Jisha2021}. Their implementation in metasurfaces has been reviewed in~\cite{Guo2022}, while nonlinear geometric phases have been considered separately in~\cite{Hu2025}. For structured optical fields, we also use total angular momentum projections, which more directly reflect the rotational symmetry of the full electromagnetic field.}

}
\blue{ A comment on terminology is mandatory at this point. 
Pancharatnam introduced the geometric phase in polarization optics through the interference of differently polarized
states~\cite{Pancharatnam1956-Generalizedtheoryof}, whereas Berry introduced geometric phase for the description of cyclic adiabatic evolution of quantum states~\cite{berry1984quantal}. Later on he established the precise relation between the two constructions~\cite{Berry1987-TheAdiabaticPhasea}. 
In the two-state polarization problem considered here, they share the same underlying geometrical structure.
The term Pancharatnam--Berry phase is commonly used in optics for this polarization geometric phase and does not relate to a third distinct effect.
Accordingly, in this context, we do not distinguish between the terms Pancharatnam phase and Pancharatnam--Berry phase, and use the term Berry phase when referring to the same construction from the more general geometrical viewpoint.}

\section{The simplest case of the Pancharatnam phase}
In this section, we describe the case of a polarized wave passing through a waveplate in some detail~\cite{Takenaka1973Jan, Sugic2021-Particle-liketopolog, Jurco1987Sep}.
\label{sec:simplest}

\subsection{Poincar\'e and Riemann sphere}
\begin{figure*}[ht!]
    \centering
    \includegraphics[width=0.79\linewidth]{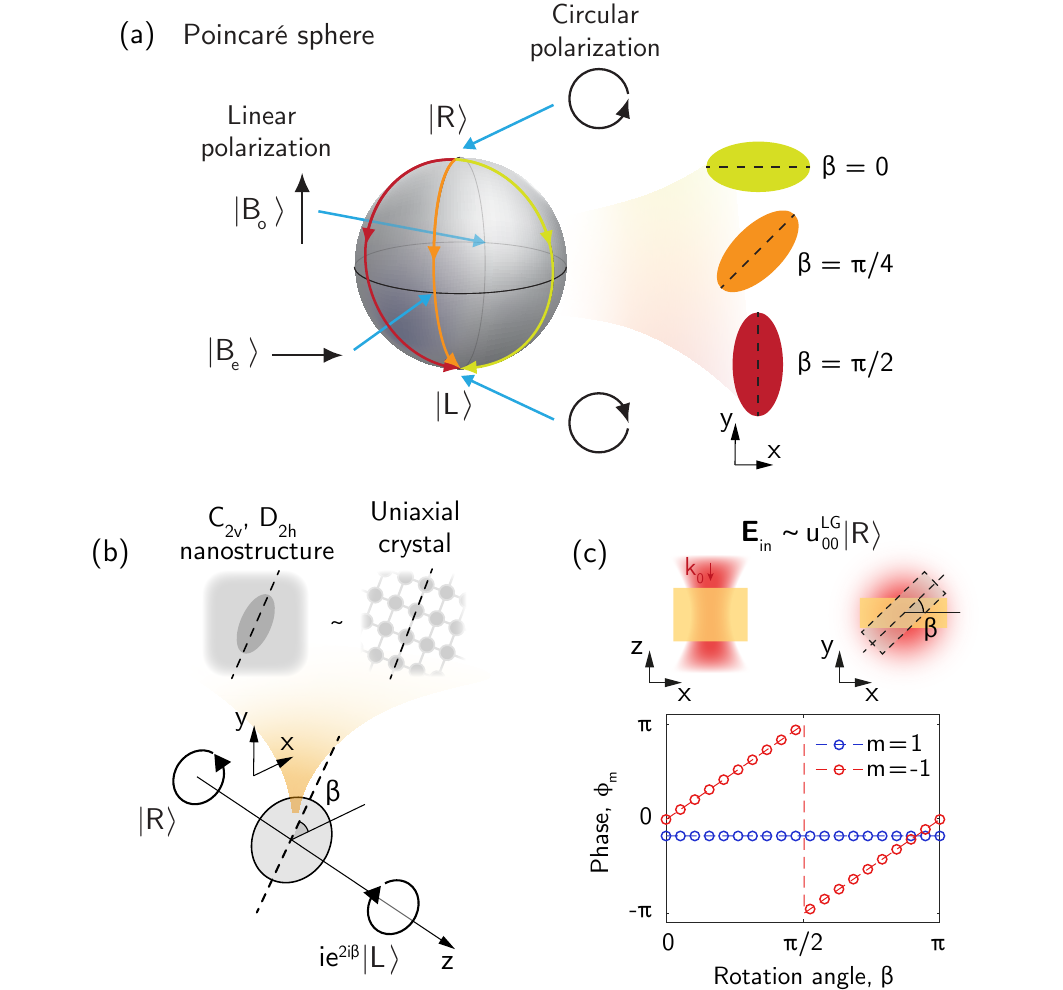}
    \caption{Poincaré sphere and examples of nanostructures for a geometric-phase experiment. 
    (a) Depending on the rotation $\beta$ of the structure, the Poincaré sphere rotates around different axes. 
    The path corresponds to the case where incident RCP light is converted to LCP light upon transmission through such a structure.
    (b) Schematic of the setup illustrating geometric phase. 
    (c) Numerically calculated phase acquired upon rotation of a rectangular nanoparticle under excitation by a fundamental Gaussian beam. \blue{The phase is wrapped to the range $[-\pi,\pi]$, circular markers indicate the calculated points, the dashed lines are guides to the eye.}
}
    \label{mainscheme}
\end{figure*}

The polarization state of light $\vec{E}$ in the basis of right $\ket{R}$ and left-circularly polarized $\ket{L}$ plane waves (RCP and LCP, respectively) can be described by two complex numbers $(E_R, E_L) \in \mathbb C^2$:
\begin{equation}
	\vec{E} = E_R \ket{R} + E_L \ket{L} =  E_x \ket{B_e} + E_y \ket{B_o}
	\label{eq:wave}
\end{equation}
where
\begin{align}
	 \ket{B_e} = \frac{\ket{R} + \ket{L}}{\sqrt{2}}, \ \ \ \
      \ket{B_o} =  -\iu\frac{\ket{R} - \ket{L}}{\sqrt{2}},
\end{align}
and $E_R$ and $E_L$ are complex numbers that fully encode the field state.
\blue{After normalization,
\begin{align}
\hat{\Psi}
=
\frac{1}{\sqrt{|E_R|^2+|E_L|^2}}
\begin{pmatrix}
E_R\\ E_L
\end{pmatrix},
\label{eq:wavenorm}
\end{align}
the pair of complex amplitudes satisfies
$|E_R|^2+|E_L|^2=1$ and therefore lies on the three-sphere
$S^3\subset\mathbb{C}^2\simeq\mathbb{R}^4$.
}
However, for determining polarization states, only the relative amplitude and phase matter; \blue{not only the overall amplitude but also} global phase factor are irrelevant.
{Let us introduce the vector:
\begin{equation}
	\vec{\Psi} = 
	\begin{pmatrix}
		E_R \\
		E_L 
	\end{pmatrix},
\end{equation}}
which is analogous to the Jones vector for linear polarization.
Since the overall amplitude and phase of $\vec{\Psi}$ do not affect the polarization state, we can normalize this vector and consider only the ratio of $E_R$ to $E_L$, represented as a complex number $z =E_R/E_L$. 
The complex plane may be compactified to a Riemann sphere, with  homogeneous coordinates $[E_R : E_L]$ (see Appendix~\ref{app:proj}).
The Poincar\'e sphere then provides a way of parametrizing polarization states in a unique fashion.
The mapping from the complex plane to the Poincar\'e sphere is given by the stereographic projection (see Appendix~\ref{app:stere} and~\cite{vid2025Nov}):
\begin{equation}
	z = \cot\left(\frac{\theta}{2}\right)e^{-\iu \varphi},
\end{equation}
where $\theta$ is the polar angle from the north pole of the Poincaré sphere and $\varphi$ is the azimuthal angle. 
This equation shows how a point on the complex plane is related to a point on the Poincaré sphere, with $\theta$ and $\varphi$ describing the polarization state.
The north and south poles of the Poincar\'e sphere ($\theta = 0$ and $\theta = \pi$, respectively) correspond to right and left circular polarizations, while points on the equator ($\theta = \pi/2$) represent linear polarizations (see Fig.~\ref{mainscheme}a).
The remaining points represent elliptical polarizations, with the azimuthal angle $\varphi$ indicating the orientation of the ellipse.

Let us consider the action of a waveplate on an arbitrarily polarized plane wave~\eqref{eq:wave}. 
Note that a similar consideration is applicable to the case depicted in Fig.~\ref{mainscheme}b, where the nanostructure with, for example $C_{2v}$, $C_{1h}$ or $D_{2h}$ symmetry plays the role of a conventional waveplate made from a uniaxial crystal. 
{From the modal perspective, this} happens because the orthogonal eigenmodes, even and odd with respect to the mirror-symmetry plane, are non-degenerate,
and excited with different phases, depending on their resonant frequencies~\cite{Gladyshev_Frizyuk_Bogdanov_2020, Xiong:20}.
First we consider the case where the waveplate is oriented with its slow axis along the $x$-direction.
As such, it just provides an additional phase $\delta$ between the components of the input Jones vector $\vec{E}=(E_x, E_y)^\top$. 
For convenience, we write the new vector as $\tilde{\vec{E}}=(E_xe^{\iu \delta/2}, E_ye^{-\iu \delta/2})^\top$.
Using the following relations
\begin{align}
E_R = \frac{E_x - \iu E_y}{\sqrt{2}}, \quad E_L = \frac{E_x + \iu E_y}{\sqrt{2}},\\
E_x = \frac{E_R + E_L}{\sqrt{2}}, \quad E_y = -\frac{E_R - E_L}{\iu\sqrt{2}},
\end{align}
we obtain new values $\tilde{E}_L$ and $\tilde{E}_R$ after the waveplate:
\begin{align} 
\label{eq:with_x}
	\begin{pmatrix}
		\tilde{E}_R \\
		\tilde{E}_L 
	\end{pmatrix}=
	\begin{pmatrix}
		\cos \delta/2 & \iu\sin \delta/2 \\
		\iu\sin \delta/2 & \cos \delta/2
	\end{pmatrix}
	\begin{pmatrix}
		E_R \\
		E_L 
	\end{pmatrix} = \nonumber 
    \\ =
	(\mathbb I \cos \delta/2 + \iu \sigma_x \sin \delta/2 )
	\begin{pmatrix}
		E_R \\
		E_L 
	\end{pmatrix}.
\end{align}
Here $\sigma_x$ is \blue{one of the Pauli} matrices
\begin{equation}    
    \sigma_x = 
        \begin{pmatrix}
            0 & 1 \\
            1 & 0
        \end{pmatrix}, \ \ 
    \sigma_y = 
        \begin{pmatrix}
            0 & -\iu \\
            \iu & 0
        \end{pmatrix}, \ \ 
    \sigma_z = 
        \begin{pmatrix}
            1 & 0 \\
            0 & -1
        \end{pmatrix}.
\end{equation}
{From \eqref{eq:with_x}, one can note that} if the phase shift corresponds to half-wave retardation, $\delta=\pi$, we have {perfect conversion between cross-polarized fields:} $E_L \rightarrow \iu E_R$, $E_R \rightarrow \iu E_L$.

Let us check what happens if we rotate the optical axis of a waveplate, so as to form \blue{an angle $\beta$} with the $x$-axis. 
We provide the derivations for this case in Appendix~\ref{app:waveplate}. 
Eventually, we get the general formula for a plane wave transmitted through a {waveplate with arbitrary retardance $\delta$}  (or a nanostructure with 2-fold rotational symmetry):
\begin{equation} 
	\begin{pmatrix}
		\tilde{E}_R \\
		\tilde{E}_L 
	\end{pmatrix}=
	(\mathbb I \cos \delta/2 + \iu (\cos 2\beta\sigma_x+\sin 2\beta\sigma_y) \sin \delta/2 )
	\begin{pmatrix}
		E_R \\
		E_L 
	\end{pmatrix}.
    \label{su2matrix}
\end{equation}

Importantly, we see that transformation~\eqref{su2matrix} is performed via an SU(2) matrix. 
One can indeed
easily check that the determinant of this matrix 
is equal to 1,
and $U^\dagger U = 1$.
For a waveplate rotated by the angle $\beta = 45^\circ$ we get
\begin{equation} 
	\begin{pmatrix}
		\tilde{E}_R \\
		\tilde{E}_L 
	\end{pmatrix}=
	(\mathbb I \cos \delta/2 + \iu \sigma_y \sin \delta/2 )
	\begin{pmatrix}
		E_R \\
		E_L 
	\end{pmatrix}.
\end{equation}
In this case for $\delta=\pi$ we have $E_R \rightarrow  -E_L$, $E_L \rightarrow E_R$, {with a different relative phase compared to the case above}. 
{From Eq.~\eqref{su2matrix} one can see that} a plane wave transmitted through a half-wave plate rotated by an angle $\beta$ is transformed as follows
\begin{equation}
\label{eq:el_er}
	\begin{pmatrix}
		\tilde{E}_R \\
		\tilde{E}_L 
	\end{pmatrix}=
	\begin{pmatrix}
		0 & \iu e^{-2 \iu \beta} \\
		\iu e^{2 \iu \beta} & 0
	\end{pmatrix}
	\begin{pmatrix}
		E_R \\
		E_L 
	\end{pmatrix} 
\end{equation}
{Let us note that in the general case, the SU(2) transformation looks as follows
\begin{equation}
	\begin{pmatrix}
		\tilde{E}_R \\
		\tilde{E}_L 
	\end{pmatrix}
	=
	\left[
	\mathbb I \cos \frac{\delta}{2}
	+
	\iu
	\left(
		n_x \sigma_x+n_y \sigma_y+n_z \sigma_z
	\right)
	\sin \frac{\delta}{2}
	\right]
	\begin{pmatrix}
		E_R \\
		E_L 
	\end{pmatrix},
	\qquad
	n_x^2+n_y^2+n_z^2=1 .
	\label{su2general}
\end{equation}}
{Interestingly, by comparing Eqs.~\eqref{su2matrix} and \eqref{su2general}, one can see that a waveplate cannot realize a rotation about the $z$-axis (generated by $\sigma_z$). Such a rotation, however, can be achieved using a gyrotropic medium~\cite{Berry1987-TheAdiabaticPhasea, Shumitskaya2024Apr, Garrison1988Jan}. } 
\blue{The group $SU(2)$ is the double cover (indeed, the universal covering group) of $SO(3)$, the group of all proper rotations in three dimensions (orthogonal transformations with unit determinant)~\cite{Zee2016-GroupTheoryinaNut, Sugic2021-Particle-liketopolog}. In the present context, these are rotations of the Poincaré sphere~\cite{Saito2023Jul, Saito2024-Quantumfieldtheory}.
}
This means that actually
two SU(2) matrices correspond to each rotation of the Poincar\'e sphere.
One can check that during any transformation of the form~\eqref{su2matrix}, all points of the Poincar\'e sphere are just rotated around some horizontal axis (the orientation of the axis depends on $\beta$). 
For this, one should keep in mind 
the relation between Stokes parameters and Jones matrices~\cite{aa638919, delCastillo2013Mar} 
and compare with Chapter IV.5 ``SU(2): Double Covering and the Spinor'' in~\cite{Zee2016-GroupTheoryinaNut}.
For convenience, we restrict our attention to the evolution of an initially circularly polarized state $\ket{R}$.
The path of this state under action of a waveplate is drawn in Fig.~\ref{mainscheme}a.
\blue{Depending on the thickness of the waveplate, the path may have different lengths, therefore, it is parametrized by the waveplate thickness.}
Using~\eqref{eq:el_er} it is easy to see that the phases of the final state $\ket{L}$ will be different, and $\beta$- dependent. 
For two different paths that cut out a slice (a closed contour) from the sphere, the phase difference will be equal to the half of the solid angle $\Omega/2$~\cite{Cohen2019-Geometricphasefrom}.

\blue{Figure~\ref{mainscheme}c shows the numerically calculated geometric phase for a rectangular nanoparticle in air ($D_{2h}$) illuminated by an RCP fundamental Gaussian beam with $m=1$ and $l=p=0$ (see also Appendix~\ref{sec:num_det}). The cross-polarized component (LCP, $m=-1$) acquires a phase factor $e^{2\iu\beta}$, so its phase winds by $2\pi$ as the particle is rotated from $\beta=0$ to $\beta=\pi$. In the Poincar\'e-sphere representation, the corresponding SU(2) rotation axis simultaneously completes one full circuit of the equator. The phase of the co-polarized component ($m=1$) is independent of $\beta$.}

{We should note that the analogy between a half-wave plate and a nanostructure is not perfect. 
In the case of a non-reflective wave plate, we have a well-defined path parametrizing the polarization state of a propagating wave, which can be depicted on the Poincar\'e sphere. In the case of a single nanostructure, it is not clear what exactly should be considered as a state, nor whether it can be parametrized by a distance. Moreover, as discussed in Refs.~\cite{Yu2026Feb, Liu2016Oct, Lalanne2017May}, the question of cross-conversion efficiency should also be taken into account. One could consider the fields inside the nanostructure and only the polarization state of the forward-propagating wave component, but this seems somewhat unnatural.
However, assuming equal amplitudes of the two excited eigenmodes, one can still formally write the same SU(2) transformation.}
\begin{figure}[ht!]
    \centering
    \includegraphics[width=\linewidth]{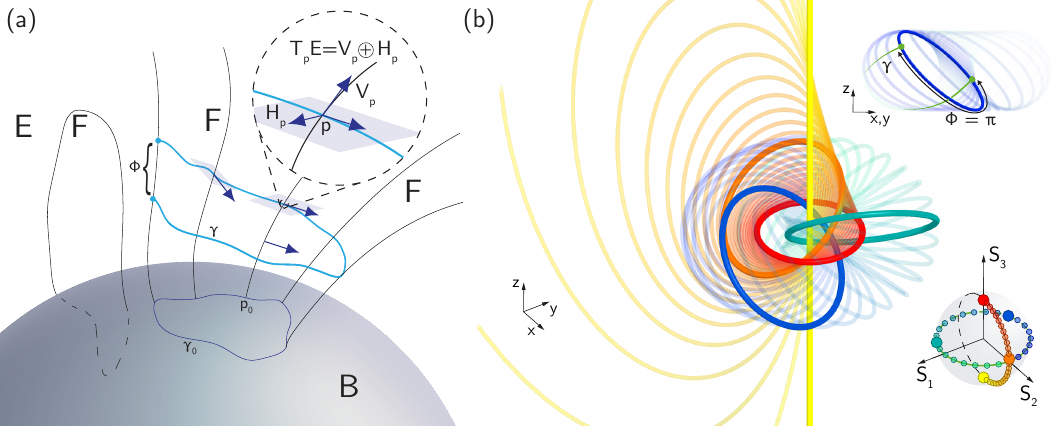}
    \caption{(a) Illustration of a circle bundle over the sphere $S^2$. 
    Fibers F are topologically circles $S^1$, the base manifold is a sphere $S^2$.
    {The horizontal lift $\gamma$ of the curve $\gamma_0$ via a connection is shown.}
    The tangent space to the total space $E$ at a point $p \in E$, $T_pE$ is shown, together with its decomposition as the direct sum of a horizontal and vertical subspace.  
    Horizontal subspaces are determined by a connection form.
    The holonomy $\Phi$ of the connection  is the geometric phase. 
    \blue{Physically, a point on the base sphere represents the polarization
    state, the fibre coordinate represents the common optical phase, and
    $\Phi$ is the phase accumulated after a closed polarization cycle.}
    {(b) Visualization of the Hopf fibration. 
    The three-dimensional sphere $S^3$ is fibered into circles and depicted in 3D space via stereographic projection. 
    The circles are mapped to polarization states on the Poincar\'e sphere $S^2$. 
    The correspondence is indicated by colors; for example, the orange circle on $S^3$ is mapped to the orange point on $S^2$. 
    The inset in \blue{the upper right} shows one of the possible horizontal lifts of the closed path along the equator of $S^2$, resulting in a holonomy of $\pi$. }}
    \label{Fig:bundle}
\end{figure}
\subsection{Differential geometric viewpoint}
Let us now consider the simple experiment discussed above 
from a differential geometric point of view. For the basic definitions, one may also consult~\cite{schuller2015lectures, Chruscinski-GeometricPhasesinC, Nakahara}.
In general, when the polarization state on the Poincar\'e sphere ($S^2$)  traverses a closed path, the actual state of light (including the common phase) traces a path in another, higher-dimensional space (see Fig.~\ref{Fig:bundle}a) which, in our specific case, is given by the three-dimensional sphere $S^3$.
\blue{This corresponds to the two normalised complex numbers in Eq.~\eqref{eq:wavenorm}}.
In this space, the path is not necessarily closed; its behavior is in fact
determined by the specific way in which the point moves, 
physically via a waveplate and mathematically through the {\it holonomy} of a connection ~\cite{Anandan1987-Somegeometricalcons, Anandan1992-Thegeometricphase, Nakahara} --- measuring the extent to which the path fails to close --- which actually represents the geometric phase. 
{The three-dimensional sphere is fibred 
into circles (corresponding to the total phase). Each circle is mapped to a single point of the base sphere $S^2$ under the bundle projection. Note, that in the Fig.~\ref{Fig:bundle}a, the fibers are shown as attached to $S^2$ only schematically, the base sphere $S^2$ should not be interpreted as an embedded submanifold of the total space $S^3$.}

This construction corresponds to the 
so-called
Hopf fibration. 
{In the same way as a two-dimensional sphere $S^2$ may be mapped onto a two-dimensional plane (Appendix~\ref{app:stere}), a three-dimensional sphere may be mapped onto a three-dimensional space. Due to the properties of stereographic projection, circles are mapped to circles or straight lines.}
{Figure~\ref{Fig:bundle}b visualizes the Hopf bundle through a stereographic projection for a set of polarization states on the Poincar\'e sphere. The colors indicate the correspondence between the fibers and the polarization states shown in the inset. Points lying at the same latitude are mapped to circles of equal radius, which form a torus. 
}
Some intuition can be also gained from specifically dedicated videos, see \textit{e.g.}~\cite{Johnson2011Aug, vid2025Nov}.
\subsubsection{Fibre bundles}
When we forget the total phase, and consider only the polarization, we have a point on a Poincar\'e sphere. 
This sphere will be called a base space. 
The total phase (we could deal with the amplitude as well, but we shall not do this here) can be depicted as a point on a circle $S^1$. 
Thus, to each point on the Poincar\'e sphere there corresponds a
circle, accommodating the total phase. 
The whole system is a \textbf{fibre bundle}, with a total space $E$, base space $B=S^2$ and fiber $F=S^1$ (see Fig.~\ref{Fig:bundle}a)~\cite{Nakahara, Chruscinski-GeometricPhasesinC}. 
We denote this arrangement as $(E, B, \pi, F)$. Here $\pi\colon E \to B$ is \blue{projection that maps} each fiber $F_x$, $x \in B$, to the point $x$ itself.
We will discuss the structure of the total space $E$ in our case in detail later on.
{So we can imagine the circles as being ``attached'' to the sphere, as the teeth of a comb are attached to its base. However, this analogy should not be understood literally: the circles are associated with the points of the sphere through the mapping $\pi$, rather than physically attached to or embedded in the sphere.}

We will work with locally trivial fibre bundles, \textit{i.e.}, those obtained by appropriately ``gluing'' 
trivial bundles $(E,U,\pi,F)$ with $E=U\times F$, where $U\subseteq B$ is an open set from an open cover of $B$ and $\pi\colon E\to U$ is the projection onto $U$; that is, for $(x,f)\in U\times F$, $\pi(x,f)=x$. 
The detailed construction of a locally trivial fibre bundle is not essential for our purposes; it suffices to know that our bundle is ``good enough''.
\subsubsection{Connections}
However, since there is no canonical way to identify the fibres with each other, we need an additional structure, a \textit{connection}, that specifies how neighbouring fibres are ``glued together''. 
For this purpose, we introduce horizontal and vertical subspaces of the tangent space  $T_pE$ to some point $p$ of the total space $E$.

The \textbf{vertical subspace} $V_p$ is the subspace of the tangent space $T_pE$ that consists of vectors that are tangent to the fibers. Formally, it is defined as:
\begin{equation}
    V_p = \{ u \in T_pE \mid \pi_* u = 0 \}, 
\end{equation} 
where $\pi\colon E \to B$ is the projection map, and $\pi_* u$ denotes the pushforward of $u$, recalled below.\par
\textbf{Pushforward}: Let $ f\colon M \to N $ be a smooth map between manifolds $ M $ and~$N$. 
\blue{The differential, or pushforward of $ f $, denoted by $ f_* $, is a map 
\begin{equation}
    f_*\colon T_pM \to T_{f(p)}N 
\end{equation} 
between the tangent spaces $ T_pM $ and $ T_{f(p)}N $, mapping a vector $ v \in T_pM $ to a vector $ f_*v \in T_{f(p)}N $.
Informally, if some vector $v$ is a tangent vector to a particular curve ${\mathcal C}$ in $M$ passing through $p$, then, after a pushforward, it becomes $f_*v$, a tangent vector at $f(p)$ to  the image curve $f({\mathcal C})$.
}
The \textbf{horizontal subspace} $H_p$ at a point $p$ in the bundle $E$ is a subspace of the tangent space $T_pE$ that complements the vertical space. 
{The term ``horizontal'' should not be understood as implying orthogonality to the vertical subspace, as no metric has been introduced so far.}
A \textbf{connection} on a bundle $(E, B, \pi, F)$ can be described in terms of a connection form $\mathcal{A}$. 
\blue{The \textbf{connection form} $\mathcal{A}$ is a 1-form on $E$ whose value specifies the vertical component of a tangent vector.}
Its action on a tangent vector $u \in T_pE$ is given by:
\blue{\begin{equation}
    \label{eq:action_on_vec}
    \mathcal{A}_p(u) \coloneq \iu a,
    \qquad
    \operatorname{ver}u = a\,u_v \in V_p,
    \quad a\in\mathbb{R}.
\end{equation}
where $\operatorname{ver}u$ denotes the vertical component of $u$,
and $u_v$ is the vertical vector field written explicitly in
Eq.~\eqref{eq:uv}.
Thus, $\mathcal{A}_p(u)$ is a purely imaginary number whose real
coefficient $a$ specifies the vertical component of $u$.}
Concretely, this 1-form just takes a vector from the tangent space  $T_pE$ and returns its projection onto the vertical subspace $V_p$, namely the component tangent to the fiber.
Equivalently, a connection specifies a complement $H_p$ to the vertical subspace $V_p$ and picks up the vertical component of a vector in $T_pE$: in other words, it is the (oblique) projection onto $V_p$ along $H_p$.
Explicitly, the
\textbf{horizontal subspace} is defined  as follows:
\[ H_p = \{ u \in T_pE \mid \mathcal{A}_p(u) = 0 \}. \]
Vertical and horizontal components are depicted in Fig.~\ref{Fig:bundle}a.
The definitions above were quite general, but for the physical problems under consideration here, we are basically interested in the celebrated Hopf bundle, to be discussed below. 
\blue{Physically, this condition fixes the choice of a common phase while the polarization changes.
For the connection used here, horizontal transport means that infinitesimally neighbouring
states are in phase in the Pancharatnam sense, see Appendix~\ref{sec:horizontal-berry-integral}.}

\subsubsection{Geometry of the Hopf bundle}
We shall illustrate the above general scheme in the important specific
case provided by the Hopf bundle, shortly ${\mathscr H}$, following ~\cite{Chruscinski-GeometricPhasesinC}.
The total space in this case is a three-dimensional sphere  $S^3$, embedded in the four-dimensional space $\mathbb C^2$, as  
\begin{equation}
\label{eq:kekek}
    |z_1|^2+|z_2|^2=1.
\end{equation} 
The base space is the two-dimensional sphere $S^2$.
The projection map $\pi \colon S^3 \to S^2$ reads 
\begin{align}
    \nonumber
   &\pi(z_1, z_2) = \\ = \nonumber & \left( 2 \operatorname{Re}(z_1 \overline{z_2}), \ 2 \operatorname{Im}(z_1 \overline{z_2}), \ |z_1|^2 - |z_2|^2 \right) = \\ & \ \ \ \ \ \ \ \ \ \ \ =   (x, y, z). \label{eq:projection}
\end{align}
These expressions for the three components resemble the Stokes parameters (with a change in notation), and this is not a coincidence.
One can check that, actually, $x^2+y^2+z^2=1$.
Note that 
\begin{align}
    \pi(z_1, z_2) = \pi(e^{\iu \alpha}z_1, e^{\iu \alpha}z_2),
\end{align}
The fibre $F$ is the circle $S^1 \approx U(1)$ (that is, we have a {\it principal bundle} with {\it structure group} $G = U(1)$) and
the above formula says that the projection is invariant under the action of $U(1)$ group.
\blue{Thus, points on the same fibre have identical polarization and differ only by the common optical phase. In other words, the point on the base sphere is determined by the Stokes
parameters, whereas motion along the $U(1)$ fibre leaves these parameters
unchanged.}

Let us now define a  ``canonical'' connection. Let \blue{ $ \langle \cdot, \cdot \rangle\colon \mathbb{C}^2 \times \mathbb{C}^2 \to \mathbb{C} $} be the standard Hermitian inner product in $ \mathbb{C}^2 $. For any $ (z_1, z_2), (w_1, w_2) \in \mathbb{C}^2 $, it is defined as:
\begin{equation}
    \label{innerproduct}
    \langle (z_1, z_2), (w_1, w_2) \rangle := \bar{z}_1 w_1 + \bar{z}_2 w_2.
\end{equation}
Let us consider a point $ p \in S^3 $ and take any vector $ v \in T_p S^3 $. 
If $ p \in S^3 $ corresponds to $ (z_1, z_2) \in \mathbb{C}^2 $, then the vertical space $ V_p $ is defined as
\begin{equation}
    \label{eq:hopf_vp}
   \blue{ V_p =
    \{(\iu a z_1, \iu a z_2)\mid a\in\mathbb R\}}
\end{equation}
For the horizontal space, we take
\begin{equation}
\label{eq:horcond}
    H_p = \{ (w_1, w_2) \in \mathbb{C}^2 \mid \langle (z_1, z_2), (w_1, w_2) \rangle = 0 \}.
\end{equation}
That is, $ H_p $ consists of all vectors tangent to $ S^3 $ at $ p $ that are orthogonal to $ p $ when $ p $ is viewed as a radius-vector in $ \mathbb{C}^2 $.
One can define the connection 1-form $ {\mathcal A} $ on $ S^3 $ as follows:
\begin{equation}
    \label{eq:connec}
    {\mathcal A} = 2g (\bar{z}_1 \dd z_1 + \bar{z}_2 \dd z_2),
\end{equation}
where $ g $ is a scaling factor. 
Indeed, action of a form on arbitrary complex vector $(v_1, v_2)$ returns its vertical component:
\begin{equation}
    2g (\bar{z}_1 \dd z_1 + \bar{z}_2 \dd z_2)(v_1, v_2) = 2g(\bar{z}_1 v_1+\bar{z}_2v_2)
\end{equation}
For the standard Hopf normalization $g = 1/2$  (compare with~\eqref{innerproduct}). 
We will keep this factor to  generalize our considerations later on.
Expanding the connection form in terms of real and imaginary parts, we get
\begin{align}
    {\mathcal A} = 2g \operatorname{Re}(\bar{z}_1 \dd z_1 + \bar{z}_2 \dd z_2) + \nonumber \\ + 2\iu g \operatorname{Im}(\bar{z}_1 \dd z_1 + \bar{z}_2 \dd z_2).
\end{align}
However, since $ (z_1, z_2) \in S^3 $, and~\eqref{eq:kekek},  we have
\begin{equation}
2 \operatorname{Re}(\bar{z}_1 \dd z_1 + \bar{z}_2 \dd z_2) = \dd (|z_1|^2 + |z_2|^2) = 0,
\end{equation}
and thus,
\begin{equation}
    {\mathcal A} = 2\iu g \operatorname{Im}(\bar{z}_1 \dd z_1 + \bar{z}_2 \dd z_2).
\end{equation}
Finally, expanding $z_1 = x_1 + \iu x_2$, and $z_2 = x_3 + \iu x_4$  we can express $ A $ as
\begin{equation}
    \label{eq:A_in4cord}
    {\mathcal A} = \blue{2\iu g} (x_1 \dd x_2 - x_2 \dd x_1 + x_3 \dd x_4 - x_4 \dd x_3),
\end{equation}
where $ x_i $ represent the real components of the vector in $ \mathbb{R}^4 $.

{Let us illustrate the action of this form in coordinates and consider the vertical and horizontal subspaces for the particular example of the Hopf fibration. 
Consider, at $p = (z_1,z_2) \in S^3$, the following orthonormal basis of tangent vectors to $S^3$:
\begin{equation}
    \label{eq:uv}
    u_v = -x_2\partial_1 + x_1\partial_2 -x_4\partial_3 +x_3\partial_4,
\end{equation}
\begin{align}
   \label{eq:uh12}
u_{h1} =  - x_3\partial_1 + x_4\partial_2 + x_1\partial_3 - x_2\partial_4,  \\
    u_{h2} =  - x_4\partial_1  - x_3\partial_2 + x_2\partial_3 + x_1\partial_4, \nonumber 
\end{align}
where a physicist can think about $\partial_i$ as unit basis vectors $\partial_i \leftrightarrow \vec e_i$ (in this particular case).
Then, for a vector $u_v$ starting at $p=(x_1,x_2,x_3,x_4)$ 
and ending at
\begin{equation}
h(t) = (x_1 - t x_2, x_2 + t x_1, x_3 - t x_4, x_4 + t x_3),
\end{equation}
\blue{we obtain
\begin{align}
    h(t) - p = (1 + \iu t)p - p = \nonumber \\ = (\iu t z_1,  \iu t z_2).
\end{align}}
which means that this vector indeed belongs to the vertical space~\eqref{eq:hopf_vp}. 
Otherwise, one can directly write the action of the 1-form~\eqref{eq:A_in4cord} on~\eqref{eq:uv}:
Using that $\dd x_j(\partial_k)=\delta_{jk}$, we have:
\begin{equation}
    \label{eq:Aonuv}
        \begin{aligned}
            {\mathcal A}(u_v)= \blue{2\iu g}\big(x_1^2 + x_2^2 + x_3^2 + x_4^2\big) = \blue{2\iu g},
    \end{aligned}
\end{equation}
since the point lies on the unit 3-sphere.  
For completeness, one can also verify that for the two other tangent vector fields~\eqref{eq:uh12}
the form gives
\begin{equation}
    {\mathcal A}(u_{h1}) = {\mathcal A}(u_{h2}) = 0,
\end{equation}
which confirms that only $u_v$ has a nonzero projection on the vertical subspace, and $u_{h1}$ and $u_{h2}$ are horizontal vector fields.
}
\subsubsection{Horizontal lift and holonomy}
In this section, we introduce the definitions that are likely the most important for understanding the Berry phase.
Let $(E, B, \pi, F)$ be a smooth fibre bundle, where $E$ is the total space ($S^3$ in our case), $B$ is the base space (Poincar\'e sphere $S^2$ in our case), $\pi\colon E \to B$ is the projection map, and $F$ is the fibre. 
Let $v \in T_xB$ be a tangent vector at $x \in B$ , and let $p \in E$  be a point such that $\pi(p) = x$. 
The \textbf{horizontal lift} of some vector $v$ at $p$ is the unique vector $\tilde{v} \in T_pE$ satisfying the following conditions:

1. Projection preservation: The vector $\tilde{v}$ projects to $v$ under the pushforward of $\pi$, \textit{i.e.},
   \begin{equation}
        \pi_*(\tilde{v}) = v,
   \end{equation}
   where $\pi_*\colon T_pE \to T_xB$. 

2. Horizontality: The vector $\tilde{v}$ lies in the horizontal subspace $H_p \subset T_pE$, defined by the connection form $\mathcal{A}$, \textit{i.e.},
   \begin{equation}
        \mathcal{A}_p(\tilde{v}) = 0.
   \end{equation}
The {\it horizontal lift $\gamma$ of a curve $\gamma_0$} on the base space is defined as follows: the projection of a lifted curve should be the initial curve, and its tangent vector should be the horizontal lift of the tangent vector to the initial curve.
In Fig.~\ref{Fig:bundle} the horizontal lift of the curve on the base space is depicted with light blue as well as lifts of tangent vectors.
In Appendix~\ref{app:lift} we provide an example in coordinates.

However, if the path on the base space is closed, the lifted path, in general, is not closed: one arrives at a different point on the fiber, as shown in the figure. This difference is called {\bf holonomy} of the connection, and corresponds exactly to the geometric phase~\cite{Simon1983-HolonomytheQuantum, Urbantke1991-xn--Twolevel-2m3dquantumsys}! An example is also given in Appendix~\ref{app:lift}.
\blue{Optically, the polarization may therefore return to its initial state while the field acquires a nonzero phase, which can be measured in relation to a reference field.}
\blue{In principle, holonomy can be introduced prior to the notion of curvature, but the latter notion is crucial for both theoretical and computational purposes.

The last definition is that of \textit{parallel transport}:
given a point $p \in E$, this is simply the motion of this point along a horizontally lifted curve.
We briefly elaborate on the above concepts in  Appendix~\ref{app:curva}.}
\subsubsection{Action of the waveplate}
\begin{figure}[h!]
    \centering
    \includegraphics[width=0.45\linewidth]{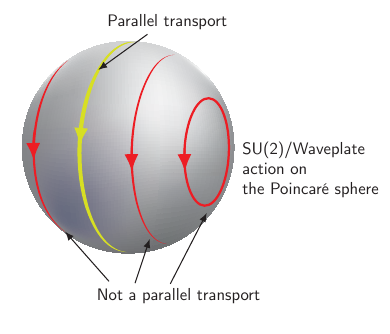}
    \caption{\blue{Illustration of the $SU(2)$ action on the Poincar\'e sphere.
A waveplate performs an $SU(2)$ transformation of the polarization state.
The condition for parallel transport is satisfied only when the state moves
along a great circle. The initially circular state considered here follows
such a trajectory.}}
    \label{Fig:paral}
\end{figure}
As we have seen earlier, the action of the waveplate is described with the help of the $SU(2)$ matrix~\eqref{su2matrix}. 
But how can we see that this action corresponds somehow to the connection form~\eqref{eq:A_in4cord}?
We discuss this in detail in the Appendix~\ref{app:rightleft}.
{The answer is the following: the motion of a point on the sphere $S^3$ under the action~\eqref{su2matrix} of an $SU(2)$ one-parameter subgroup comes from parallel transport  if and only if the corresponding projection on the Poincar\'e sphere moves along a great circle,
namely, a geodesic on $S^2$~\cite{Berry1987-TheAdiabaticPhasea, Voitiv2023-Experimentalmeasurem, Hagen:24, Courtial1999Dec}. {We illustrate this schematically in Fig.~\ref{Fig:paral}.}
}

\section{A direct Pancharatnam--Berry-phase analogy for vortex beams via the Poincar\'e sphere}
\label{sec:mainres}

{Let us now consider vortex beams.
In cylindrical coordinates, vortex beams allow for the construction of a higher-order Poincaré sphere~\cite{Padgett1999-Poincare-sphereequiv, 
Shen2021-RayswavesSU2sy}, which is fully analogous to the usual one:
\begin{align}
    |R_m\rangle 
&\sim e^{\iu m \varphi}(\hat{\bm \rho}+\iu \hat{{\bm \varphi}}),\\
|L_m\rangle 
&\sim e^{-\iu m \varphi}(\hat{\bm \rho}-\iu \hat{{\bm \varphi}}),
    \label{eq_ll_rl}
\end{align}
where $|R_m\rangle$ and $|L_m\rangle$ signify right- and left-handed vortex beams, respectively, with $m \in \mathbb Z$ denoting the TAM projection, $\varphi$ the azimuthal angle, and $\hat{\bm \rho}$, $\hat{{\bm \varphi}}$ are unit vectors of cylindrical coordinate system.
The equivalence sign $\sim$ stands for the symmetry behavior, \textit{i.e.} behavior under rotations around the $z$-axis. 
\begin{figure*}[ht!]
    \centering
    \includegraphics[width=0.79\linewidth]{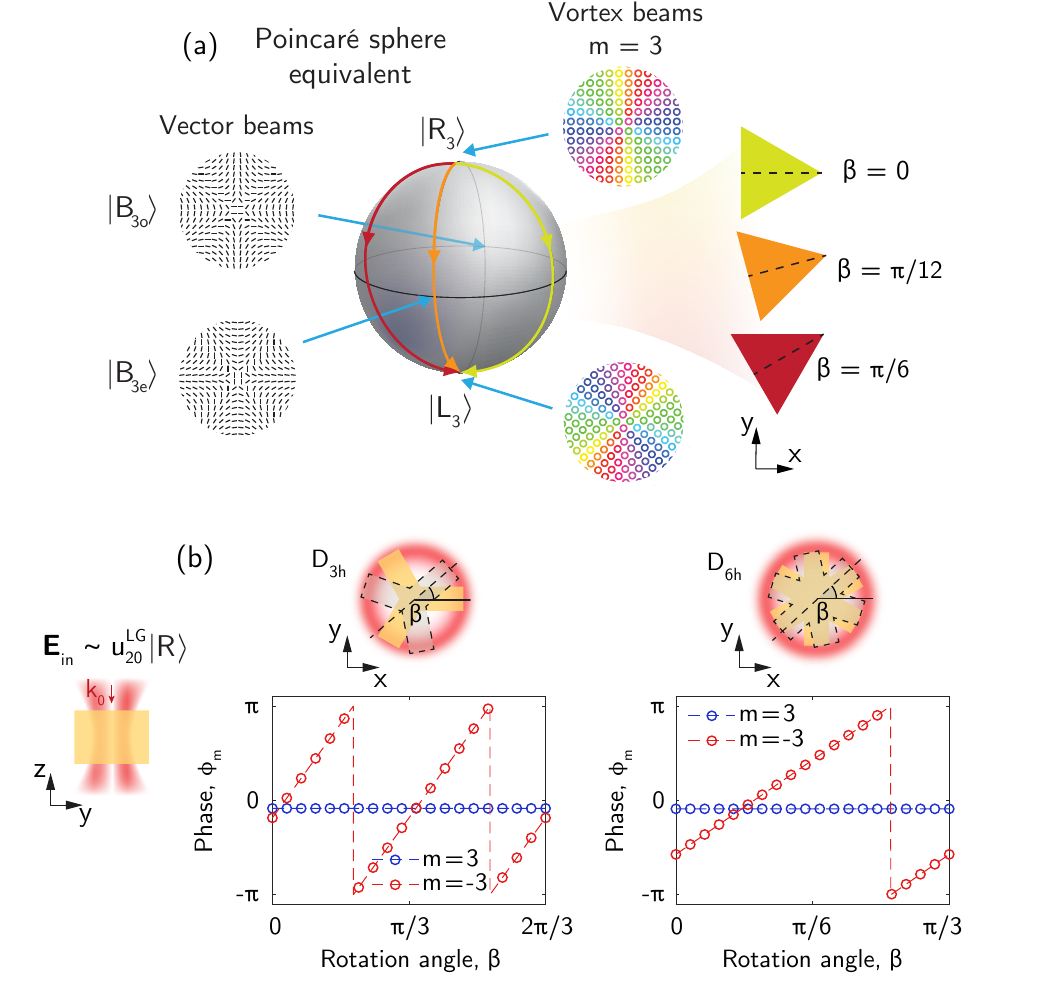}
    \caption{(a) Geometry of the experiment involving the transmission of vortex beams with TAM projection $m$ through a prism with $\mathfrak{n}$-fold rotational symmetry.
This setup is fully analogous to the one depicted in~Fig.~\ref{mainscheme}, but different $\beta$ is required.
    {(b)~Numerically calculated phase acquired by the rotation of nanoparticles with $D_{3h}$ and $D_{6h}$ symmetries under excitation by a vortex beam ($m$ = 3). \blue{The phase is wrapped to the range $[-\pi,\pi]$, circular markers indicate the calculated points, the dashed lines are guides to the eye. For both nanoparticles the condition $2m=\mathfrak{n}\nu$ is satisfied, therefore 2$\pi$ phase is accumulated by $\pi/3$ rotation.}}
    }
    \label{mainscheme2}
\end{figure*}
{Precisely, this means that under rotation of the vector function as a whole by some angle $\alpha$ around the $z$-axis, the function does not change except for an additional phase $\text{e}^{\pm \iu m \alpha}$.
See \textit{e.g.}~\cite{Tung2020-GroupTheoryInPhysi, Menshikov2025Nov} for a precise definition; a detailed discussion can be found in~\cite{Nikitina2024-AchiralNanostructure}.}
Note, that 
\begin{equation}
    (\hat{\vec x}\pm \iu \hat{{\vec y}}) = e^{\pm \iu \varphi}(\hat{\bm \rho}\pm \iu \hat{{\bm \varphi}}),
    \label{eq:cyl_dec}
\end{equation}
and the sign refers to the polarization (helicity) here.
We omit the dependence on the radial coordinate~$r$~\cite{Zhan2006-Propertiesofcircula}, since it does not alter the general symmetry we are interested in.}
Similarly {to Eq.~\ref{eq:wave} we can introduce the following linear combination, or a vector vortex beam:}
\begin{equation}
    \vec E = c_1  |R_m\rangle+ c_2  |L_m\rangle.
    \label{eq_ll2}
\end{equation}
At this stage we want to describe an experiment, what would be analogous to the one with the waveplate.
{A {vector} vortex beam is fundamentally different from a plane wave, being anisotropic in the transverse direction.} 
\blue{{As already discussed,} for plane waves, not only waveplates but also geometric-phase metasurfaces with rectangular dielectric metaatoms are widely used~\cite{Song2020-Ptychographyretrieva, Overvig2019-Dielectricmetasurfac}. Related PB elements based on subwavelength dielectric gratings and patterned liquid crystals were demonstrated in~\cite{Hasman2003-Polarizationdependen, Marrucci2006May}, see also the review~\cite{Cohen2019-Geometricphasefrom}.}

Such metasurfaces, as well as liquid-crystal-based setups governed by the same principles, are often used for structured-light generation~\cite{Marrucci2006May, Larocque2016Nov}.
Indeed, each metaatom has, \textit{e.g.,} $C_{2v}$, $D_{2h}$ or $C_{1h}$ symmetry. 
For these symmetries, two eigenmodes, one of which is excited \blue{by an $x$-polarized wave and the other by} an $y$-polarized wave, are transformed according to the different irreducible representations, and, as a consequence, have different resonance frequencies and are excited with different phases~\cite{Gladyshev_Frizyuk_Bogdanov_2020, Nikitina2024-AchiralNanostructure}. 
This consideration via the eigenmodes and irreps may appear too complicated for this simple case of a rectangular block; however, it can be easily adapted to the vortex beam case.
{We expand the discussion on the modal excitation in the Appendix~\ref{app:modes}.}
Indeed, vector beams, which are defined as:
\begin{align}
    |\text{B}_{me}&\rangle \propto |L_m\rangle + |R_m\rangle\\
|\text{B}_{mo}&\rangle \propto |L_m\rangle - |R_m\rangle,
    \label{eq:vb_o_e}
\end{align}
where the indexes $e$ and $o$ denote the parity under reflection in the $ y=0 $ plane, refer to the endpoints of a diameter of the higher-order Poincar\'e sphere.
They serve as an analogy for vertical and horizontal linear polarization in the plane wave case.
Thus, instead of the rectangular block, one should use the nanostructure, in which $ |\text{B}_{me}\rangle$ and $ |\text{B}_{mo}\rangle$ are transformed according to the different irreps. 
This metaatom should have $C_{\mathfrak{n}v}$ or $D_{\mathfrak{n}h}$ symmetry with $2m=\mathfrak{n}\nu, \ \ \nu \in \mathbb Z$ (See also~\cite{Nikitina2024-AchiralNanostructure}, Section 8).
For such a structure, these vector beams will excite eigenmodes which have different phases, leading to exactly the same process previously discussed. 
Note that this process is not perfect because other values of $m$ are also obtained during scattering. 
Since one cannot avoid them, the only way out is to tailor the geometry so as to minimize their contribution. 
\blue{Although this process is analogous to the simplest case, it presents only one difference. }
In order to obtain the same phase difference $\delta$ one should rotate the nanostructure by a smaller angle $\beta$ (see Fig.~\ref{mainscheme}a). 
One can also notice
that opposite points on the diameter refer to the beam rotation by $\beta=\pi/(2m)$, as well as the nanostructure rotation for the opposite paths on the Poincare sphere.
For a plane wave $m=\pm 1$.
\blue{It is also analogous to the transformation implemented by a q-plate, as described in~\cite{SuzukiPhysRevA.94.043851}. 
Although the physical implementations are different, the detailed shape of the q-plate is not essential for this analogy.}

Let us also note that, in principle, if we have an incident wave with TAM projection $m^{\text{inc}}$ and an output wave of the same \blue{frequency with the TAM} projection $m^\text{out}$, the acquired phase will be equal to~\cite{guercio2026tensordrivengeometricphasenonlinear}
\begin{equation}
\label{eq:phasem}
    \phi(\beta) =( m^{\text{inc}} - m^\text{out})\beta,
\end{equation}
however, the corresponding geometric interpretation is not straightforward for other $|m^{\text{out}}| \neq | m^\text{inc}|$.
Moreover,  in view of the selection rules for scattering, 
only $m^{\text{out}} = m^\text{inc} + \mathfrak{n}\mathbb{Z}$ are allowed.

{Figure \ref{mainscheme2}b shows a numerical calculation of geometric phase for particles of different symmetries illuminated by a beam with $m=3$ featuring second order phase singularity (LG mode with $l=2$, $p=0$, see Appendix~\ref{sec:num_det}). One can see that both $D_{3h}$ and $D_{6h}$ particles with the same input, provide a $2\pi$ phase winding of the cross polarized component under rotation by the angle $\pi/3$.}
\blue{In summary, the construction is the direct analogue of the
polarization case, with the two opposite-TAM states replacing the two
circular-polarization states and the rotation of the nanostructure
controlling the acquired phase.}

\section{The usual geometric phase of vortex beams via the Hermite--Gaussian sphere}
\label{sec:HG_beams}
Let us now proceed to another experiment, which is commonly associated with the geometric phase of optical beams. 
This experiment possesses a completely different geometry, when compared to the one described above.
First, we consider only the scalar part of the beam, because polarization may be considered independently 
for this experiment.
Here we will use the paraxial approximation. 
In the previous case, no such approximation was needed.
The Hermite-Gaussian modes are solutions of the paraxial wave equation~\cite{Lax1975-FromMaxwelltoparax, Nienhuis1993-Paraxialwaveoptics, Kogelnik1966-LaserBeamsandReson} 
\begin{equation}
\frac{\partial^2}{\partial x^2} u(\vec r)+\frac{\partial^2}{\partial y^2} u(\vec r) = -2\iu k \frac{\partial}{\partial z} u(\vec r),
\end{equation}
and may be written as
\begin{equation}
    u_{\nu\eta}^\text{HG}(\vec{r}) = \frac{1}{\gamma} \psi_\nu \left( \frac{x}{\gamma} \right) \psi_\eta \left( \frac{y}{\gamma} \right) \exp \left[ \frac{\iu k r_t^2}{2R} - \iu  \chi (\nu + \eta + 1) \right],
\end{equation}   
where $\psi_n(\xi) $ are the normalized Hermite functions corresponding to the one-dimensional harmonic oscillator:
\begin{equation}
    \psi_n(\xi) = \frac{1}{\sqrt{2^n n! \sqrt{\pi}}} \, e^{ -\xi^2/2} H_n(\xi).
\end{equation}
$\chi=\chi(z)$ is the Gouy phase, $\gamma=\gamma(z)$ is the spot radius and $R=R(z)$ is the radius of the curvature of the wavefront, $\eta, \nu \in  \mathbb{N}_0$, $k$ is a wavenumber, $r_t^2=x^2+y^2$, $H_n(\xi)$ is a Hermite polynomial~\cite{Nienhuis1993-Paraxialwaveoptics}.
The solutions can be also presented in a different form:
\begin{equation}
    u_{\ell p}^\text{LG}(\mathbf{r}) = \exp \left[ \frac{ik}{2R}r_t^2 - \iu\chi(N + 1) \right] \frac{1}{\gamma} \psi_{lp} 
    \left(\frac{r_t}{\gamma}, \varphi\right),
    \label{eq:lg_expr}
\end{equation}
\blue{
\begin{equation}
    \psi_{\ell p}(\rho,\varphi)=\sqrt{\frac{p!}{\pi(p+|\ell|)!}} \, e^{ -\rho^2/ 2}\, e^{\iu \ell \varphi }\, \rho^{|\ell|} \, L_p^{|\ell|}(\rho^2),
    \label{eq:lg_expr2}
\end{equation}
}
where $N=\eta+\nu=2p+|\ell|$ and $L_p^{|\ell|}$ is the generalized Laguerre polynomial. 
Notice that there are $N+1$ solutions for each $N$.
For our purposes, an exact form is not necessary, because we are only interested in the following presentation~\cite{Nienhuis1993-Paraxialwaveoptics, Nienhuis2004-Angularmomentumand, Beijersbergen1993-Astigmaticlasermode}:
\begin{equation}
    u_{\nu\eta}^\text{HG}(\vec r) = \frac{1}{\sqrt{\nu! \eta!}} \left( \hat{a}_x^\dagger \right)^\nu \left( \hat{a}_y^\dagger \right)^\eta u_{00}(\vec r)
\end{equation}
where 
\begin{equation}
    \hat{a}_x = \frac{1}{\sqrt{2bk}} \left[ kx + (b + \iu z) \frac{\partial}{\partial x} \right],
\end{equation}
\begin{equation}
    \hat{a}_x^\dagger = \frac{1}{\sqrt{2bk}} \left[ kx - (b - \iu z) \frac{\partial}{\partial x} \right].
\end{equation}
and $\hat{a}_y$ is defined analogously with $x \to y$~\cite{Nienhuis1993-Paraxialwaveoptics, Nienhuis2004-Angularmomentumand}, $b$ is the Rayleigh range,  $\chi(z)=\arctan(z/b)$, and $\frac{1}{b+\iu z}
    = \frac{1}{k \gamma^2} -  \frac{\iu}{R}$~\cite{Nienhuis1993-Paraxialwaveoptics}.
Let us define the operators~\cite{Dennis2017-Swingsandroundabout, Calvo2005-Wignerrepresentation, Simon2000-Wignerrepresentation}
\begin{align}
    L_x =& \frac{1}{2} (\hat{a}_x^\dagger \hat{a}_x - \hat{a}_y^\dagger \hat{a}_y) 
    =\frac{1}{4bk}\!\left[k^2(x^2-y^2) + 2\iu kz\,(x\partial_x-y\partial_y) -(b^2+z^2)\,(\partial_x^2-\partial_y^2)\right]. \\
    L_y =& \frac{1}{2} (\hat{a}_x^\dagger \hat{a}_y + \hat{a}_y^\dagger \hat{a}_x)    
    =\frac{1}{2bk}\!\left[ k^2xy + \iu kz\,(x\partial_y+y\partial_x) -(b^2+z^2)\,\partial_x\partial_y\right] \\
    L_z =& \frac{1}{2\iu} (\hat{a}_x^\dagger \hat{a}_y - \hat{a}_y^\dagger \hat{a}_x) 
     =\frac{\iu}{2}\,(y\partial_x-x\partial_y) \\
    L_0 =& \frac{1}{2} (\hat{a}_x^\dagger \hat{a}_x + \hat{a}_y^\dagger \hat{a}_y) = \\ 
    = &\frac{1}{4bk}\!\left[ k^2(x^2+y^2) +2\iu kz\,(x\partial_x+y\partial_y) -2k(b-\iu z) -(b^2+z^2)\,(\partial_x^2+\partial_y^2) \right] \nonumber.
\end{align}
\blue{For fixed mode order $N$, $L_x$, $L_y$, and $L_z$ generate SU(2)
transformations within the transverse-mode subspace, analogous to polarization
rotations on the Poincar\'e sphere.}
For $z=0$ we have~\cite{Simon2000-Wignerrepresentation} 
\begin{equation}
    L_x u_{\nu\eta}^{HG}(\mathbf{r}) = \frac{1}{2} (\nu - \eta) u_{\nu\eta}^{HG}(\mathbf{r}),
\end{equation}
and introducing
\begin{equation}
    \hat{a}^\dagger_{\pm} = \frac{\hat{a}^\dagger_x \pm \iu \hat{a}^\dagger_y}{\sqrt{2}},
\end{equation}
we get
\begin{equation}
    \label{eq:lp_crea}
    u_{\ell p}^\text{LG}(\vec r) = \frac{1}{\sqrt{\nu_+! \nu_-!}} \left(\hat{a}^\dagger_+\right)^{\nu_+} \left(\hat{a}^\dagger_-\right)^{\nu_-} u_{00}(\vec r),
\end{equation}
\begin{equation}
    L_z u_{\ell p}^{LG}(\mathbf{r}) = \frac{1}{2} (\nu_+ - \nu_-) u_{\ell p}^{LG}(\mathbf{r}).
\end{equation}
where $\ell =\nu_+-\nu_-, \ p=\min(\nu_+, \nu_-)$ \blue{and $N =\nu_++\nu_-$}.

Let us imagine an experiment, which would change the ``relative phase'' $\delta$ between $ \hat{a}^\dagger_x$ and $ \hat{a}^\dagger_y$.
This would mean, that $u_{\nu\eta}^{\text{HG}}(\vec r)$  will acquire
some additional relative phase equal to $(\nu-\eta) \delta/2$. 
This can be realized with the mode converter (\cite{Beijersbergen1993-Astigmaticlasermode}, eq. (20)-(22), \cite{Allen1999-Matrixformulationfo}).
\blue{In an astigmatic mode converter, this relative phase is produced by the
different Gouy-phase shifts accumulated along the two transverse axes.}
Thus, 
\begin{equation}
    u_{\ell p}^\text{LG}(\vec r) = \frac{1}{\sqrt{\nu_+! \nu_-!}}  \frac{\left(\hat{a}^\dagger_x + \iu \hat{a}^\dagger_y\right)}{\sqrt{2}}^{\nu_+}  \frac{\left(\hat{a}^\dagger_x - \iu \hat{a}^\dagger_y\right)}{\sqrt{2}}^{\nu_-} u_{00}(\vec r),
\end{equation}
transforms into
\begin{equation}
    u_{\ell p}^{\prime\text{LG}}(\vec r) = \frac{1}{\sqrt{\nu_+! \nu_-!}}  \frac{\left(e^{\iu\delta/2}\hat{a}^\dagger_x + \iu e^{-\iu\delta/2}\hat{a}^\dagger_y\right)}{\sqrt{2}}^{\nu_+}  \frac{\left(e^{\iu\delta/2}\hat{a}^\dagger_x - \iu e^{-\iu\delta/2}\hat{a}^\dagger_y\right)}{\sqrt{2}}^{\nu_-} u_{00}(\vec r),
    \label{eq:lg_phas}
\end{equation}

Let us note, that in each multiplier, $\hat{a}^\dagger_i$ behaves just in the same way 
as $E_i$ in previous waveplate considerations.
Thus, one can depict all the states on the Hermite-Laguerre sphere, which would be a Poincar\'e-like sphere, constructed for creation operators.
One can also consider a fiber bundle associated with such a sphere. 
Note that it is important, that all the multipliers behave identically. 
On the one side, just by comparison 
with a half-waveplate, one may consider the 3-sphere $S^3$, associated with complex numbers $(c_1, c_2)$, associated with ``state'' $\Phi$, which describes the phases of the creation operators:
\begin{align}
    &\Phi = c_1\hat{a}^\dagger_+ + c_2\hat{a}^\dagger_- \\
    &\blue{\Phi_\perp = -c_2^*\hat{a}^\dagger_+ + c_1^*\hat{a}^\dagger_-, }\\
    &|c_1|^2 + |c_2|^2 = 1.
\end{align}
\blue{For Eq.~\eqref{eq:lg_phas}, analogously to Eq.~\eqref{eq:with_x}, the corresponding coefficients are
\begin{equation}
    c_1=\cos\frac{\delta}{2},
    \qquad
    c_2=\iu\sin\frac{\delta}{2},
\end{equation}
{defined up to an arbitrary common phase factor $e^{\iu\alpha}$}. The actual state of a beam Eq.~\eqref{eq:lg_phas} can be written simply as
\begin{equation}
    u_{\ell p}^{\prime\text{LG}}(\vec r)
    =
    \frac{1}{\sqrt{\nu_+!\nu_-!}}
    \Phi^{\nu_+}\Phi_\perp^{\nu_-}u_{00}(\vec r).
    \label{eq:lg_phas_ren}
\end{equation}}
For the case of two operators, the geometry is just the same as in the simplest case. 
However, one should keep in mind that each point on the operator 2-sphere 
would correspond to some state of the beam~\eqref{eq:lg_phas}, which we want to be a point in the total space, so the fiber coordinate
corresponds to the phase of the beam.
{From \eqref{eq:lg_phas_ren} one can see that}, the common phase turns out to be a phase $(\nu_+-\nu_-)\alpha= l \alpha$, where $\alpha$ corresponds to the ``phase'' of the operators $\Phi$. 
\blue{For $\ell\neq0$, points of a fibre whose phases $\alpha$ differ by $2\pi/|\ell|$
should be considered equivalent,  giving rise to a smooth orbifold, namely the lens space $L_{|\ell|,1}$~\cite{Albers2023-ASymplecticDynamics, BibEntry2024Nov}. This circle bundle has first Chern number $\ell$, corresponding to the monopole strength $g=\ell/2$~\cite{Bruzzo2023-D3-branesupergravity}.}
{The connection form has a similar form, but with multiplier $g=\ell/2$ in \eqref{eq:connec}.}
Polarization is considered independently, and modified, as before, via the waveplate.
If polarization follows the same rotation path, the total phase may be expressed in terms of the TAM projection $m=\ell+\sigma$.
\blue{Thus, the essential difference from the polarization case lies not in the
fibre coordinate itself, which again represents the optical phase, but in the
base space and in how its evolution gives rise to the accumulated phase.}

{For higher-order modes, which consist of, \textit{e.g.}, three basis states, an octant representation of the base space may be useful~\cite{Cisowski2026Feb}. 
In this case, the particular geometry of the whole bundle remains an open question.}

\subsection{Majorana representation}
Let us write again the relations between different indices, and introduce $J$ and $\mu$: 
\blue{
\begin{align}
    J=N/2,\ \quad \mu &= \ell/2,\\
    J+\mu = \frac{N+\ell}{2}&=\nu_+,\\
    J-\mu = \frac{N-\ell}{2}&=\nu_-,
\end{align}
}
and express the state
\eqref{eq:lp_crea} as follows~\cite{Lee2022-TopologyandGeometry, Schwinger2015-OnAngularMomentum, Shabbir2017-MajoranaRepresentati, Majorana1932-Atomiorientatiinca, Gutierrez-Cuevas2020-ModalMajoranaSphere}:
\begin{align}
    \left| J, \mu \right\rangle = \frac{(a_+^\dagger)^{J+ \mu}}{\sqrt{(J+ \mu)!}} \frac{(a_-^\dagger)^{J- \mu}}{\sqrt{(J- \mu)!}}\ket{0, 0}.
\end{align}
Let us consider an arbitrary beam \blue{ of fixed order $N$}:

\begin{align}
    \left| \Psi \right\rangle &= \sum_{\mu=-J}^{J} c_\mu \left| J,\mu \right\rangle = \sum_{\mu=-J}^{J} c_\mu \frac{(a_+^\dagger)^{J+\mu} (a_-^\dagger)^{J-\mu}}{\sqrt{(J+\mu)!(J-\mu)!}} \ket{0, 0},
    \label{eq:majo}
\end{align}
or, equivalently
\begin{align}
    \left| \Psi \right\rangle =  \sum_{\mu=-J}^{J} c_\mu \frac{(a_+^\dagger)^{J+\mu} (a_-^\dagger)^{-(J+\mu)}}{\sqrt{(J+\mu)!(J-\mu)!}} (a_-^\dagger)^{2J} \ket{0, 0}.
    \label{eq:majo2}
\end{align}
\blue{For $c_J\neq 0$ let $-z_i$} be the roots of the polynomial of a power $2J$  in the variable $x$ (formally one can consider $ (a_+^\dagger) (a_-^\dagger)^{-1}$ as $x$ in~\eqref{eq:majo2}) and with index shift $J+\mu = i$:
\begin{align}
     \sum_{i=0}^{2J} c_{i-J}\frac{x^i}{\sqrt{(2J-i)!i!}}=0 \leftrightarrow  \frac{c_J}{\sqrt{(2J)!}}\prod_{i=1}^{2J}(x+z_i) = 0.
    \label{eq:majo45}
\end{align}
Keeping in mind the fundamental theorem of algebra, one can also write~\eqref{eq:majo2} in the same form:
\begin{align}
    \left| \Psi \right\rangle = c_J \sqrt{\frac{1}{(2J)!}} \prod_{i=1}^{2J} (a_+^\dagger + z_i a_-^\dagger) \ket{0, 0}.
    \label{eq:maj2}
\end{align}
\blue{If $c_J=0$, the polynomial in~\eqref{eq:majo45} has degree lower than $2J$. Its finite roots are supplemented by roots at infinity so that their total number remains $2J$. In the factorized state~\eqref{eq:maj2}, each root at infinity corresponds to a factor $a_-^\dagger$.}
Thus, we have a correspondence between coefficients $c_\mu$ and roots $z_i$, which can be depicted on the sphere $S^2$ with the help of stereographic projection.
For a \blue{Laguerre-Gaussian} beam only one $c_{\mu_0}$ is non-zero, thus we have $J+\mu_0$ roots $z_i=0$, and $J-\mu_0$ at infinity.
The action of the mode converter rotates the roots collectively.
\blue{\section{A zoo of Poincar\'e spheres}
\label{sec:poincare_zoo}


There are many different objects called Poincar\'e spheres in the structured-light literature. 
Most of them are constructed in the same way: one chooses two orthogonal basis states and considers their linear combinations as in~\eqref{eq:wave}. 
As discussed in Section~\ref{sec:simplest} and Appendix~\ref{app:proj}, for two states $\ket{N}$ and $\ket{S}$ we can write
\begin{equation}
    \ket{\Psi(\theta,\varphi)}
    =
    \cos\frac{\theta}{2}\ket{N}
    +
    e^{\iu\varphi}
    \sin\frac{\theta}{2}\ket{S},
    \qquad
    0\leq\theta\leq\pi,
    \quad
    0\leq\varphi<2\pi.
    \label{eq:poincare_zoo_state}
\end{equation}
Thus, in order to define a particular Poincar\'e-like sphere, one has to specify its two pole states. 
In the context of structured light and paraxial approximation, we characterize these states by their polarization and topological charge. 

One of the first spatial-mode Poincar\'e spheres was introduced by Padgett and Courtial for first-order modes~\cite{Padgett1999-Poincare-sphereequiv}. In their construction, the two poles correspond to the LG modes with $\ell=+1$ and $\ell=-1$. 
The same formal construction can be extended to arbitrary $\ell$:
\begin{equation}
    \ket{N}=\ket{\pi}u^{\mathrm{LG}}_{+\ell,p},
    \qquad
    \ket{S}=\ket{\pi}u^{\mathrm{LG}}_{-\ell,p},
    \label{eq:higher_order_orbital_sphere}
\end{equation}
where $\ket{\pi}$ is a fixed polarization state. Thus, polarization does not change on this sphere. This construction is often called the \emph{orbital Poincar\'e sphere}~\cite{Perumangatt2015Generalized, Karimi2010PolarizationControlled, Yang2025May}.

The term ``orbital'' is potentially misleading. In paraxial optics, this index has historically been referred to as an orbital angular momentum. However, $\ell$ is topological charge and does not define the orbital angular momentum of an individual photon, since the spin and orbital parts of the angular momentum of a massless field are not separable~\cite{Palmerduca2025-Fournogotheorems,Fernandez-Corbaton2012-Helicityandangular}. Thus, \emph{orbital Poincar\'e sphere} is just a historical name. 
Moreover, the same name is also used by some authors~\cite{Calvo2005-Wignerrepresentation, Rao2026Jun} in a different sense, for a different construction described below.

For $\ell=1$, the transformation of the state on the sphere Eq.~\eqref{eq:higher_order_orbital_sphere} is produced by the standard astigmatic mode converter~\cite{Lassen2009-ContinuousVariableEntanglement}. 
This is not true for higher orders. 
For example, one can define a sphere with $u^{\mathrm{LG}}_{+2,0}$ and $u^{\mathrm{LG}}_{-2,0}$ at its poles, but the intermediate states on this sphere are not the states produced by a cylindrical-lens converter~\cite{Habraken2010-Universaldescription,Habraken2010-Geometricphasesinh}. 
 The mode converter instead transforms states on the \emph{Hermite--Laguerre sphere}~\cite{Habraken2010-Geometricphasesinh} discussed also in Section~\ref{sec:HG_beams}. 
 This sphere is constructed for the creation operators $\hat a^\dagger_\pm$ introduced in Eq.~\eqref{eq:lp_crea}, while their transformation is given by Eq.~\eqref{eq:lg_phas}. For $N=1$, the \emph{Hermite--Laguerre sphere} coincides with the \emph{orbital Poincar\'e sphere} in Eq.~\eqref{eq:higher_order_orbital_sphere}. For $N>1$, powers of the transformed operators produce additional cross terms and, consequently, additional modes of the same order. Therefore, the two constructions may have the same pole modes but different states between the poles. 

\emph{Higher-order Poincar\'e spheres} are obtained by using both  polarization and phase vortex~\cite{Holleczek2011May}. From Eq.~\eqref{eq:cyl_dec},
\begin{equation}
    \ket{N}
    \sim
    \ket{R}\,u^{\mathrm{LG}}_{m-1,p},
    \qquad
    \ket{S}
    \sim
    \ket{L}\,u^{\mathrm{LG}}_{1-m,p}.
    \label{eq:zoo_tam_charge_relation}
\end{equation}
The states $\ket{N}$ and $\ket{S}$ have TAM projections $+m$ and $-m$, and topological charges $m-1$ and $1-m$, respectively.
This is the higher-order Poincar\'e sphere considered in Section~\ref{sec:mainres}. Holleczek \textit{et al.} considered two such spaces for first-order cylindrically polarized fields~\cite{Holleczek2011May} with the difference in the choice of the polar axis. 

The \emph{hybrid-order Poincar\'e sphere} of Yi \textit{et al.} is constructed in the same way, but the topological charges of its two poles are independent~\cite{Yi2015-Hybrid-orderPoincare}:
\begin{equation}
    \ket{N}
    \sim
        \ket{R}\,u^{\mathrm{LG}}_{\ell_1,p},
            \qquad
    \ket{S}
    \sim
        \ket{L}\,u^{\mathrm{LG}}_{\ell_2,p}
    \label{eq:hyops_zoo}
\end{equation}
The two polarizations are orthogonal, but $\ell_1$ and $\ell_2$ do not have to be opposite. For $\ell_2=-\ell_1$, Eq.~\eqref{eq:hyops_zoo} reduces to the higher-order Poincar\'e sphere above.

The sphere used by Ling \textit{et al.} for full Poincar\'e beams is a particular hybrid-order Poincar\'e sphere~\cite{Ling2016-Characterizationandmanipulation}:
\begin{equation}
    \ket{N}
    \sim
        \ket{R}\,u^{\mathrm{LG}}_{0,0},
                    \qquad
    \ket{S}
    \sim
        \ket{L}\,u^{\mathrm{LG}}_{\ell,0}
    \label{eq:full_poincare_zoo}.
\end{equation}

The \emph{generalized Poincar\'e sphere} of Ren \textit{et al.} combines a continuous family of such two-state spheres~\cite{Ren2015-GeneralizedPoincare}. For each fixed value of the parameter $R_{\mathrm G}$, the corresponding space can be written as
\begin{equation}
    \ket{N}
    \sim
        \ket{\pi(R_{\mathrm G})}\,
        u^{\mathrm{LG}}_{-\ell,p},
        \qquad
    \ket{S}
    \sim
        \ket{\pi^\perp(R_{\mathrm G})}\,
        u^{\mathrm{LG}}_{\ell,p},
    \qquad
    \left\langle
        \pi^\perp(R_{\mathrm G})
        \middle|
        \pi(R_{\mathrm G})
    \right\rangle
    =0.
    \label{eq:generalized_poincare_zoo}
\end{equation}
The parameter $R_{\mathrm G}\in[1/2,1]$ is represented as the radius of the generalized sphere. A surface with a fixed radius $R_{\mathrm G}$ is called a spherical shell. Moving from one shell to another changes the ellipticity of the pole polarizations themselves. 
The radial coordinate  labels which orthogonal polarization basis is used to construct the sphere. 

Thus, most objects in this ``zoo'' differ only in the two states used as their basis. The Hermite--Laguerre sphere is different: it is constructed for creation operators.
Finally, for three or more basis states, the full projective state space is no longer a sphere, Poincar\'e spheres describe only two-state subspaces. 
Higher-dimensional states require other representations, such as the octant constructions~\cite{Cisowski2026Feb}.}

\section{Discussion}
\label{sec:discussion}
We wish to point out
that in the literature the term \textit{geometric phase} is not always used consistently. 
In particular, while for a wave plate the trajectory on the Poincaré sphere is clearly defined and can be parametrized by a real coordinate, this is no longer the case for the corresponding nanoparticle. 
In a nanoparticle, the polarization cannot be always defined or tracked
inside the structure in the same way as in a waveplate, and the associated unitary matrices are therefore introduced in a rather formal sense. 
Nevertheless, for nanoparticles, the phase can be obtained by a different procedure, namely by rotating the particle and moving to a different reference frame.

For the so-called nonlinear geometric phase, there exists an analogy with this approach~\cite{Tymchenko2015-GradientNonlinearPa}, however, in this case it is apparently no longer possible, even formally, to introduce a unitary matrix or a closed path~\cite{suzuki2016commenthigherorderpancharatnamberry} on the Poincar\'e sphere. 
This raises the question of whether all these phases should indeed be termed geometric, or whether the concept of geometric phase should be extended to a broader class of problems. 
We regard the strict delineation of this class of phases in optics as an open question for discussion.

\begin{figure}[ht!]
    \centering
    \includegraphics[width=0.5\linewidth]{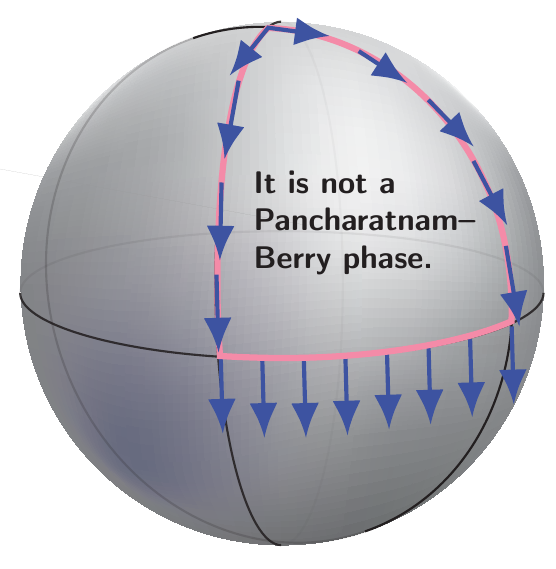}
    \caption{A frequently used illustration of a geometric phase which, however, is not a Pancharatnam-Berry phase and has a different geometry; it is equal to the full solid angle rather than half of it.}
    \label{notberry}
\end{figure}

We further suggest to reserve the term \textit{Pancharatnam-Berry phase} exclusively for phases whose geometry coincides with that described here and is associated with the Hopf fibration. 
Other phases, for example, the frequently used illustration (see Fig.~\ref{notberry}) based on parallel transport of a tangent vector on a sphere are indeed geometric phases, but they can be misleading when used in discussions of the Pancharatnam–Berry phase, as they are characterized by a fundamentally different geometry.
{This picture based on parallel transport of a tangent vector on \(S^2\)
corresponds not to the Hopf fibration, but to the unit tangent bundle with a different value of Chern class $c_1(TS^2)=2$, which is also the Euler class, whereas the Hopf bundle $S^3\to S^2$ has
$c_1=1$~\cite{HatcherVBKT, DavisKirkAlgebraicTopology}.
Its total space is $L(2,1)\simeq SO(3)  \simeq \mathbb{RP}^3$,
rather than \(S^3\). 
Thus the two constructions define genuinely different circle bundles
over the base $S^2$, even though both have fibre $S^1$.}
{

Table~\ref{tab:gp_summary} summarizes the main geometrical settings discussed above. 
The same rotation-dependent phase factor does not by itself define a unique geometric phase; the relevant distinction is the underlying base space and fibre structure.

\begin{table*}[t]
\centering
\small
\setlength{\tabcolsep}{3pt}
\renewcommand{\arraystretch}{1.25}
\caption{Summary of the geometrical settings discussed in the tutorial.}
\label{tab:gp_summary}
\begin{tabularx}{\textwidth}{
>{\raggedright\arraybackslash}p{0.18\textwidth}
>{\raggedright\arraybackslash}p{0.21\textwidth}
>{\raggedright\arraybackslash}p{0.25\textwidth}
>{\raggedright\arraybackslash}p{0.21\textwidth}
>{\raggedright\arraybackslash}p{0.08\textwidth}}
\toprule
Case & Base space & Total space & Phase law & PB? \\
\midrule

Polarization waveplate &
Poincar\'e sphere &
Hopf bundle &
$\Omega/2$, $e^{\pm 2i\beta}$ &
Yes \\

Nanostructure, $2m_{\rm out/in}=\mathfrak{n}\nu$ &
Formal Poincar\'e analogy &
Formal Hopf bundle &
$\Omega/2$, $e^{\iu (m_{\rm inc}-m_{\rm out})\beta}$ &
Restricted \\

Arbitrary TAM pair &
Formal two-state sphere &
Hopf bundle after choosing the two-state subspace &
$e^{\iu (m_{\rm inc}-m_{\rm out})\beta}$ &
Weak analogy \\

HG/LG usual consideration &
Hermite--Laguerre sphere  &
Lens spaces &
${\ell \Omega}/2$ &
Not PB \\

Arrows on sphere, unit tangent bundle &
Sphere $S^2$ &
$L(2,1)\simeq SO(3)  \simeq \mathbb{RP}^3$ &
$\Omega$ &
Not PB \\

\bottomrule
\end{tabularx}
\end{table*}
}

{Another open issue concerns the non-uniqueness of the connection. 
The connection discussed here is the canonical one, however, it is in principle possible to introduce other connections that are not related to it by a gauge transformation and that yield different curvature and different holonomy, while still corresponding to the same Chern class. 
To our knowledge, such situations have not been discussed in the optics literature. Nevertheless, it cannot be excluded that for other physical problems precisely these alternative connections may turn out to be relevant.
}
{It might then be expedient to resort to the classical complex algebraic-geometric formalism employed in~\cite{SANSONETTO2010501, PENNA199899}.
On the experimental side, some of the theoretical issues discussed in~\cite{Barbierigeometry} may be also fruitfully explored.}

\section{Conclusions}
Summing up, in this paper
we have analyzed the geometric (Pancharatnam) phase, starting from its abstract mathematical definition and establishing an explicit bridge to physical observables through concrete calculations.
We have further demonstrated that for vortex beams, an entirely analogous Berry phase can be introduced, which is equal to $\Omega/2 $, rather than $ \ell \Omega/2$, within a distinct experimental configuration. 
\blue{The comparison shows that similar phase laws may arise from different
underlying geometries, which can be important for the interpretation and design
of structured-light transformations. This viewpoint may be useful for
mode-selective nanophotonic devices and optical mode multiplexing, while its
extension to more general multimode and nonlinear systems remains an open
direction.}
\blue{As new types of optical geometric phases continue to emerge, it becomes
increasingly important to move beyond heuristic analogies and specify their
geometry precisely. Similar phase laws or Poincar\'e-like representations do not
by themselves imply the same underlying mechanism. A formulation in terms of the
relevant state space, bundle, and connection provides a common framework in which
different geometric phases can be defined, compared, and consistently classified.}

\section{Acknowledgments}
K.F. thanks Timur Seidov, Tim Sulimov, Igor Shenderovich, Nikita Belousov, Costantino de Angelis for the fruitful discussions.  Part of this work was conducted at the University of Brescia.
K.F. gratefully acknowledges support from the Alexander von Humboldt Foundation. \par
{E.M. acknowledges the PNRR RESTART project Smart Metasurfaces Advancing Radio Technology (SMART), CUP E63C22002040007, and Erasmus Mundus EMIMEP, CUP D81I24000080006.} \par
{ M.S. thanks Gabriele Barbieri for enlightening discussions and collaboration on related topics.
His research is supported by UCSC D1-funds and it has been carried out within INDAM-GNSAGA's framework.}

\blue{The authors acknowledge the use of ChatGPT (OpenAI) for assistance with language editing and improving the clarity of the manuscript.
}
\bibliographystyle{iopart-num}
\bibliography{sample}

\providecommand{\newblock}{}
\begin{thebibliography}{100}
\expandafter\ifx\csname url\endcsname\relax
  \def\url#1{{\tt #1}}\fi
\expandafter\ifx\csname urlprefix\endcsname\relax\def\urlprefix{URL }\fi
\providecommand{\eprint}[2][]{\url{#2}}

\bibitem{Cisowski2022-Colloquium}
Cisowski C, G{\ifmmode\ddot{o}\else\"{o}\fi}tte J~B and Franke-Arnold S 2022
  {\em Rev. Mod. Phys.\/} {\bf 94} 031001 ISSN 1539-0756
  \urlprefix\url{https://doi.org/10.1103/RevModPhys.94.031001}

\bibitem{Cohen2019-Geometricphasefrom}
Cohen E, Larocque H, Bouchard F, Nejadsattari F, Gefen Y and Karimi E 2019 {\em
  Nat. Rev. Phys.\/} {\bf 1} 437--449 ISSN 2522-5820
  \urlprefix\url{https://doi.org/10.1038/s42254-019-0071-1}

\bibitem{Pancharatnam1956-Generalizedtheoryof}
Pancharatnam S 1956 {\em Proc. Indian Acad. Sci.\/} {\bf 44} 247--262 ISSN
  0370-0089 \urlprefix\url{https://doi.org/10.1007/BF03046050}

\bibitem{Berry1987-TheAdiabaticPhasea}
Berry M 1987 {\em Journal of Modern Optics\/} {\bf 34} 1401--1407
  \urlprefix\url{https://doi.org/10.1080/09500348714551321}

\bibitem{Ding2015-UltrathinPancharatna}
Ding X, Monticone F, Zhang K, Zhang L, Gao D, Burokur S~N, de~Lustrac A, Wu Q,
  Qiu C~W and Al{\ifmmode\grave{u}\else\`{u}\fi} A 2015 {\em Adv. Mater.\/}
  {\bf 27} 1195--1200 ISSN 0935-9648
  \urlprefix\url{https://doi.org/10.1002/adma.201405047}

\bibitem{Luo2017-TransmissiveUltrathi}
Luo W, Sun S, Xu H~X, He Q and Zhou L 2017 {\em Phys. Rev. Appl.\/} {\bf 7}
  044033 ISSN 2331-7019
  \urlprefix\url{https://doi.org/10.1103/PhysRevApplied.7.044033}

\bibitem{Huang2012-DispersionlessPhase}
Huang L, Chen X, M{\ifmmode\ddot{u}\else\"{u}\fi}hlenbernd H, Li G, Bai B, Tan
  Q, Jin G, Zentgraf T and Zhang S 2012 {\em Nano Lett.\/} {\bf 12} 5750--5755
  ISSN 1530-6984 \urlprefix\url{https://doi.org/10.1021/nl303031j}

\bibitem{doi:10.1126/science.aaf6644}
Khorasaninejad M, Chen W~T, Devlin R~C, Oh J, Zhu A~Y and Capasso F 2016 {\em
  Science\/} {\bf 352} 1190--1194
  \urlprefix\url{https://www.science.org/doi/abs/10.1126/science.aaf6644}

\bibitem{Wen2015-HelicityMultiplexed}
Wen D, Yue F, Li G, Zheng G, Chan K, Chen S, Chen M, Li K~F, Wong P~W~H, Cheah
  K~W, Pun E~Y~B, Zhang S and Chen X 2015 {\em Nature Communications\/} {\bf 6}
  8241 \urlprefix\url{https://doi.org/10.1038/ncomms9241}

\bibitem{Zheng2015-MetasurfaceHolograms}
Zheng G, M{\"u}hlenbernd H, Kenney M, Li G, Zentgraf T and Zhang S 2015 {\em
  Nature Nanotechnology\/} {\bf 10} 308--312
  \urlprefix\url{https://www.nature.com/articles/nnano.2015.2}

\bibitem{Marrucci2006May}
Marrucci L, Manzo C and Paparo D 2006 {\em Appl. Phys. Lett.\/} {\bf 88} 221102
  ISSN 0003-6951 \urlprefix\url{https://doi.org/10.1063/1.2207993}

\bibitem{Xie2021-GeneralizedPancharat}
Xie X, Pu M, Jin J, Xu M, Guo Y, Li X, Gao P, Ma X and Luo X 2021 {\em Phys.
  Rev. Lett.\/} {\bf 126} 183902 ISSN 1079-7114
  \urlprefix\url{https://doi.org/10.1103/PhysRevLett.126.183902}

\bibitem{DENG2023106730}
Deng Q, Yang J, Lan X, Zhang W, Cui H, Xie Z, Li L and Huang Y 2023 {\em
  Results in Physics\/} {\bf 51} 106730 ISSN 2211-3797
  \urlprefix\url{https://www.sciencedirect.com/science/article/pii/S2211379723005235}

\bibitem{Galvez2003-GeometricPhaseAssoc}
Galvez E~J, Crawford P~R, Sztul H~I, Pysher M~J, Haglin P~J and Williams R~E
  2003 {\em Phys. Rev. Lett.\/} {\bf 90} 203901 ISSN 1079-7114
  \urlprefix\url{https://doi.org/10.1103/PhysRevLett.90.203901}

\bibitem{Padgett1999-Poincare-sphereequiv}
Padgett M~J and Courtial J 1999 {\em Opt. Lett.\/} {\bf 24} 430--432
  \urlprefix\url{https://opg.optica.org/ol/abstract.cfm?URI=ol-24-7-430}

\bibitem{vanEnk1993-Geometricphasetran}
van Enk S~J 1993 {\em Opt. Commun.\/} {\bf 102} 59--64 ISSN 0030-4018
  \urlprefix\url{https://doi.org/10.1016/0030-4018(93)90472-H}

\bibitem{Habraken2010-Universaldescription}
Habraken S~J~M and Nienhuis G 2010 {\em Opt. Lett.\/} {\bf 35} 3535--3537 ISSN
  1539-4794 \urlprefix\url{https://doi.org/10.1364/OL.35.003535}

\bibitem{Habraken2010-Geometricphasesinh}
Habraken S~J~M and Nienhuis G 2010 {Geometric phases in higher-order transverse
  optical modes} {\em {Proceedings Volume 7613, Complex Light and Optical
  Forces IV}\/} vol 7613 (SPIE) pp 121--128
  \urlprefix\url{https://doi.org/10.1117/12.840024}

\bibitem{Tomita1986}
Tomita A and Chiao R~Y 1986 {\em Phys. Rev. Lett.\/} {\bf 57} 937--940 ISSN
  0031-9007 \urlprefix\url{https://doi.org/10.1103/PhysRevLett.57.937}

\bibitem{Bliokh2008Geometrodynamics}
Bliokh K~Y, Niv A, Kleiner V and Hasman E 2008 {\em Nat. Photonics\/} {\bf 2}
  748--753 ISSN 1749-4885
  \urlprefix\url{https://doi.org/10.1038/nphoton.2008.229}

\bibitem{Zhen2023}
Zhen W, Wang X~L, Ding J and Wang H~T 2023 {\em Phys. Rev. A\/} {\bf 108}
  023514 ISSN 2469-9926
  \urlprefix\url{https://doi.org/10.1103/PhysRevA.108.023514}

\bibitem{Zhen2025}
Zhen W, Ren Z~C, Wang X~L, Ding J and Wang H~T 2025 {\em Sci. China Phys. Mech.
  Astron.\/} {\bf 68} 244211 ISSN 1674-7348
  \urlprefix\url{https://doi.org/10.1007/s11433-024-2589-6}

\bibitem{Wang2023Mobius}
Wang J, Valligatla S, Yin Y, Schwarz L, Medina-S{\'a}nchez M, Baunack S, Lee
  C~H, Thomale R, Li S, Fomin V~M, Ma L and Schmidt O~G 2023 {\em Nat.
  Photonics\/} {\bf 17} 120--125 ISSN 1749-4885
  \urlprefix\url{https://doi.org/10.1038/s41566-022-01107-7}

\bibitem{Palmerduca2024Apr}
Palmerduca E and Qin H 2024 {\em Phys. Rev. D\/} {\bf 109} 085005
  \urlprefix\url{https://doi.org/10.1103/PhysRevD.109.085005}

\bibitem{Cao2024}
Cao Z~X, Liu T~T, Zhao B, Rosales-Guzm{\'a}n C, Liu J and Zhu Z~H 2024 {\em
  Opt. Express\/} {\bf 32} 28025--28034 ISSN 1094-4087
  \urlprefix\url{https://doi.org/10.1364/OE.530909}

\bibitem{Simon1983-HolonomytheQuantum}
Simon B 1983 {\em Phys. Rev. Lett.\/} {\bf 51} 2167--2170 ISSN 1079-7114
  \urlprefix\url{https://doi.org/10.1103/PhysRevLett.51.2167}

\bibitem{Cayssol2021-Topologicalandgeome}
Cayssol J and Fuchs J~N 2021 {\em J. Phys.: Mater.\/} {\bf 4} 034007 ISSN
  2515-7639 \urlprefix\url{https://doi.org/10.1088/2515-7639/abf0b5}

\bibitem{Rao2026}
Rao A~S 2026 {\em Adv. Photonics Res.\/} {\bf 7} e70220 ISSN 2699-9293
  \urlprefix\url{https://doi.org/10.1002/adpr.70220}

\bibitem{Jisha2021}
Jisha C~P, Nolte S and Alberucci A 2021 {\em Laser Photonics Rev.\/} {\bf 15}
  2100003 ISSN 1863-8899 \urlprefix\url{https://doi.org/10.1002/lpor.202100003}

\bibitem{Guo2022}
Guo Y, Pu M, Zhang F, Xu M, Li X, Ma X and Luo X 2022 {\em Photonics
  Insights\/} {\bf 1} R03 ISSN 2791-1748
  \urlprefix\url{https://doi.org/10.3788/PI.2022.R03}

\bibitem{Hu2025}
Hu Z and Li G 2025 {\em Appl. Phys. Lett.\/} {\bf 126} 100502 ISSN 0003-6951
  \urlprefix\url{https://doi.org/10.1063/5.0252617}

\bibitem{berry1984quantal}
Berry M~V 1984 {\em Proceedings of the Royal Society of London. A. Mathematical
  and Physical Sciences\/} {\bf 392} 45--57

\bibitem{Takenaka1973Jan}
Takenaka H 1973 {\em Nouv. Rev. Opt.\/} {\bf 4} 37 ISSN 0335-7368
  \urlprefix\url{https://doi.org/10.1088/0335-7368/4/1/304}

\bibitem{Sugic2021-Particle-liketopolog}
Sugic D, Droop R, Otte E, Ehrmanntraut D, Nori F, Ruostekoski J, Denz C and
  Dennis M~R 2021 {\em Nat. Commun.\/} {\bf 12} 1--10 ISSN 2041-1723
  \urlprefix\url{https://doi.org/10.1038/s41467-021-26171-5}

\bibitem{Jurco1987Sep}
Jur{\ifmmode\check{c}\else\v{c}\fi}o B 1987 {\em Czech. J. Phys.\/} {\bf 37}
  1035--1038 ISSN 1572-9486 \urlprefix\url{https://doi.org/10.1007/BF01597447}

\bibitem{vid2025Nov}
{Vektorfeld} 2025 {From Poincar{\ifmmode\acute{e}\else\'{e}\fi} sphere (Bloch
  sphere) to Projective space}
  \urlprefix\url{https://www.youtube.com/watch?v=Qun3gahLP3o}

\bibitem{Gladyshev_Frizyuk_Bogdanov_2020}
Gladyshev S, Frizyuk K and Bogdanov A 2020 {\em Phys. Rev. B\/} {\bf 102}
  075103 ISSN 2469-9969
  \urlprefix\url{https://doi.org/10.1103/PhysRevB.102.075103}

\bibitem{Xiong:20}
Xiong Z, Yang Q, Chen W, Wang Z, Xu J, Liu W and Chen Y 2020 {\em Opt.
  Express\/} {\bf 28} 3073--3085
  \urlprefix\url{https://opg.optica.org/oe/abstract.cfm?URI=oe-28-3-3073}

\bibitem{Shumitskaya2024Apr}
Shumitskaya A~A, Kozlov V~O, Selivanov N~I, Stoumpos C~C, Zapasskii V~S,
  Kapitonov Y~V and Ryzhov I~I 2024 {\em Adv. Opt. Mater.\/} {\bf 12} 2302095
  ISSN 2195-1071 \urlprefix\url{https://doi.org/10.1002/adom.202302095}

\bibitem{Garrison1988Jan}
Garrison J~C and Chiao R~Y 1988 {\em Phys. Rev. Lett.\/} {\bf 60} 165--168
  \urlprefix\url{https://doi.org/10.1103/PhysRevLett.60.165}

\bibitem{Zee2016-GroupTheoryinaNut}
Zee A 2016 {\em {Group Theory in a Nutshell for Physicists}\/} (Princeton, NJ,
  USA: Princeton University Press) ISBN 978-0-69116269-0
  \urlprefix\url{https://www.amazon.com/Group-Theory-Nutshell-Physicists-Zee/dp/0691162697}

\bibitem{Saito2023Jul}
Saito S 2023 {\em Front. Phys.\/} {\bf 11} 1225419 ISSN 2296-424X
  \urlprefix\url{https://doi.org/10.3389/fphy.2023.1225419}

\bibitem{Saito2024-Quantumfieldtheory}
Saito S 2024 {\em Front. Phys.\/} {\bf 11} 1225334 ISSN 2296-424X
  \urlprefix\url{https://doi.org/10.3389/fphy.2023.1225334}

\bibitem{aa638919}
Mechanic Q 2021 {\em Physics Stack Exchange\/} (\textit{Preprint}
  \eprint{https://physics.stackexchange.com/q/638919})
  \urlprefix\url{https://physics.stackexchange.com/q/638919}

\bibitem{delCastillo2013Mar}
del Castillo G~F~T and Rubalcava-Garcia I 2013 {\em arXiv\/} (\textit{Preprint}
  \eprint{1303.4496}) \urlprefix\url{https://doi.org/10.48550/arXiv.1303.4496}

\bibitem{Yu2026Feb}
Yu L, Singh H~J, Pietila J and Caglayan H 2026 {\em Light Sci. Appl.\/} {\bf
  15} 119 ISSN 2047-7538
  \urlprefix\url{https://doi.org/10.1038/s41377-025-02153-w}

\bibitem{Liu2016Oct}
Liu C, Bai Y, Zhao Q, Yang Y, Chen H, Zhou J and Qiao L 2016 {\em Sci. Rep.\/}
  {\bf 6} 34819 ISSN 2045-2322
  \urlprefix\url{https://doi.org/10.1038/srep34819}

\bibitem{Lalanne2017May}
Lalanne P and Chavel P 2017 {\em Laser Photonics Rev.\/} {\bf 11} 1600295 ISSN
  1863-8880 \urlprefix\url{https://doi.org/10.1002/lpor.201600295}

\bibitem{schuller2015lectures}
Schuller F~P 2015 {\em Institute for Quantum Gravity, Friedrich-Alexander
  Universit{\"a}t Erlangen-N{\"u}rnberg\/}

\bibitem{Chruscinski-GeometricPhasesinC}
Chru{\ifmmode\acute{s}\else\'{s}\fi}ci{\ifmmode\acute{n}\else\'{n}\fi}ski D and
  Jamio{\l}kowski A 2004 {\em Geometric Phases in Classical and Quantum
  Mechanics\/} (Boston, MA, USA: Birkh{\ifmmode\ddot{a}\else\"{a}\fi}user) ISBN
  978-0-8176-8176-0
  \urlprefix\url{https://link.springer.com/book/10.1007/978-0-8176-8176-0}

\bibitem{Nakahara}
Nakahara M 2003 {\em Geometry, Topology and Physics, Second Edition (Graduate
  Student Series in Physics)\/} (Boca Raton, FL, USA: CRC Press) ISBN
  978-0-75030606-5

\bibitem{Anandan1987-Somegeometricalcons}
Anandan J and Stodolsky L 1987 {\em Phys. Rev. D\/} {\bf 35} 2597--2600 ISSN
  2470-0029 \urlprefix\url{https://doi.org/10.1103/PhysRevD.35.2597}

\bibitem{Anandan1992-Thegeometricphase}
Anandan J 1992 {\em Nature\/} {\bf 360} 307--313 ISSN 1476-4687
  \urlprefix\url{https://doi.org/10.1038/360307a0}

\bibitem{Johnson2011Aug}
Johnson N 2011 {Hopf fibration -- fibers and base}
  \urlprefix\url{https://www.youtube.com/watch?v=AKotMPGFJYk}

\bibitem{Urbantke1991-xn--Twolevel-2m3dquantumsys}
Urbantke H 1991 {\em Am. J. Phys.\/} {\bf 59} 503--509 ISSN 0002-9505
  \urlprefix\url{https://doi.org/10.1119/1.16809}

\bibitem{Voitiv2023-Experimentalmeasurem}
Voitiv A~A, Lusk M~T and Siemens M~E 2023 {\em Opt. Lett.\/} {\bf 48}
  2680--2683 ISSN 1539-4794 \urlprefix\url{https://doi.org/10.1364/OL.489899}

\bibitem{Hagen:24}
Hagen N and Garza-Soto L 2024 {\em J. Opt. Soc. Am. A\/} {\bf 41} 2014--2022
  \urlprefix\url{https://doi.org/10.1364/JOSAA.538106}

\bibitem{Courtial1999Dec}
Courtial J 1999 {\em Opt. Commun.\/} {\bf 171} 179--183 ISSN 0030-4018
  \urlprefix\url{https://doi.org/10.1016/S0030-4018(99)00473-3}

\bibitem{Shen2021-RayswavesSU2sy}
Shen Y 2021 {\em J. Opt.\/} {\bf 23} 124004 ISSN 2040-8986
  \urlprefix\url{https://doi.org/10.1088/2040-8986/ac3676}

\bibitem{Tung2020-GroupTheoryInPhysi}
Tung W~K 2020 {\em {Group Theory In Physics: An Introduction To Symmetry
  Principles, Group Representations, And Special Functions In Classical And
  Quantum Physics}\/} (London, England, UK: World Scientific) ISBN
  978-0-00098975-8
  \urlprefix\url{https://www.amazon.com/Group-Theory-Physics-Introduction-Representations/dp/0000989754}

\bibitem{Menshikov2025Nov}
Menshikov E, Franceschini P, Frizyuk K, Fernandez-Corbaton I, Tognazzi A, Cino
  A~C, Garoli D, Petrov M, de~Ceglia D and De~Angelis C 2025 {\em Light Sci.
  Appl.\/} {\bf 14} 381 ISSN 2047-7538
  \urlprefix\url{https://doi.org/10.1038/s41377-025-02004-8}

\bibitem{Nikitina2024-AchiralNanostructure}
Nikitina A and Frizyuk K 2024 {\em Advanced Optical Materials\/} {\bf 12}
  2400732
  \urlprefix\url{https://advanced.onlinelibrary.wiley.com/doi/abs/10.1002/adom.202400732}

\bibitem{Zhan2006-Propertiesofcircula}
Zhan Q 2006 {\em Opt. Lett.\/} {\bf 31} 867--869 ISSN 1539-4794
  \urlprefix\url{https://doi.org/10.1364/OL.31.000867}

\bibitem{Song2020-Ptychographyretrieva}
Song Q, Baroni A, Sawant R, Ni P, Brandli V, Chenot S,
  V{\ifmmode\acute{e}\else\'{e}\fi}zian S, Damilano B, de~Mierry P, Khadir S,
  Ferrand P and Genevet P 2020 {\em Nat. Commun.\/} {\bf 11} 1--8 ISSN
  2041-1723 \urlprefix\url{https://doi.org/10.1038/s41467-020-16437-9}

\bibitem{Overvig2019-Dielectricmetasurfac}
Overvig A~C, Shrestha S, Malek S~C, Lu M, Stein A, Zheng C and Yu N 2019 {\em
  Light Sci. Appl.\/} {\bf 8} 1--12 ISSN 2047-7538
  \urlprefix\url{https://doi.org/10.1038/s41377-019-0201-7}

\bibitem{Hasman2003-Polarizationdependen}
Hasman E, Kleiner V, Biener G and Niv A 2003 {\em Appl. Phys. Lett.\/} {\bf 82}
  328--330 ISSN 0003-6951 \urlprefix\url{https://doi.org/10.1063/1.1539300}

\bibitem{Larocque2016Nov}
Larocque H, Gagnon-Bischoff J, Bouchard F, Fickler R, Upham J, Boyd R~W and
  Karimi E 2016 {\em J. Opt.\/} {\bf 18} 124002 ISSN 2040-8986
  \urlprefix\url{https://doi.org/10.1088/2040-8978/18/12/124002}

\bibitem{SuzukiPhysRevA.94.043851}
Suzuki M, Yamane K, Oka K, Toda Y and Morita R 2016 {\em Phys. Rev. A\/} {\bf
  94}(4) 043851
  \urlprefix\url{https://link.aps.org/doi/10.1103/PhysRevA.94.043851}

\bibitem{guercio2026tensordrivengeometricphasenonlinear}
Guercio G, Gerini A, Frizyuk K, Morassi M, Lema{\^i}tre A, De~Angelis C,
  Carletti L and Leo G 2026 {\em ACS Photonics\/} {\bf 13} 3134--3140 ISSN
  2330-4022 \urlprefix\url{https://doi.org/10.1021/acsphotonics.6c00299}

\bibitem{Lax1975-FromMaxwelltoparax}
Lax M, Louisell W~H and McKnight W~B 1975 {\em Phys. Rev. A\/} {\bf 11}
  1365--1370 ISSN 2469-9934
  \urlprefix\url{https://doi.org/10.1103/PhysRevA.11.1365}

\bibitem{Nienhuis1993-Paraxialwaveoptics}
Nienhuis G and Allen L 1993 {\em Phys. Rev. A\/} {\bf 48} 656--665 ISSN
  2469-9934 \urlprefix\url{https://doi.org/10.1103/PhysRevA.48.656}

\bibitem{Kogelnik1966-LaserBeamsandReson}
Kogelnik H and Li T 1966 {\em Appl. Opt.\/} {\bf 5} 1550--1567 ISSN 2155-3165
  \urlprefix\url{https://doi.org/10.1364/AO.5.001550}

\bibitem{Nienhuis2004-Angularmomentumand}
Nienhuis G and Visser J 2004 {\em J. Opt. A: Pure Appl. Opt.\/} {\bf 6} S248
  ISSN 1464-4258 \urlprefix\url{https://doi.org/10.1088/1464-4258/6/5/020}

\bibitem{Beijersbergen1993-Astigmaticlasermode}
Beijersbergen M~W, Allen L, van~der Veen H~E~L~O and Woerdman J~P 1993 {\em
  Opt. Commun.\/} {\bf 96} 123--132 ISSN 0030-4018
  \urlprefix\url{https://doi.org/10.1016/0030-4018(93)90535-D}

\bibitem{Dennis2017-Swingsandroundabout}
Dennis M~R and Alonso M~A 2017 {\em Philos. Trans. Royal Soc. A\/} {\bf 375}
  ISSN 1471-2962 \urlprefix\url{https://doi.org/10.1098/rsta.2015.0441}

\bibitem{Calvo2005-Wignerrepresentation}
Calvo G~F 2005 {\em Opt. Lett.\/} {\bf 30} 1207--1209 ISSN 1539-4794
  \urlprefix\url{https://doi.org/10.1364/OL.30.001207}

\bibitem{Simon2000-Wignerrepresentation}
Simon R and Agarwal G~S 2000 {\em Opt. Lett.\/} {\bf 25} 1313--1315 ISSN
  1539-4794 \urlprefix\url{https://doi.org/10.1364/OL.25.001313}

\bibitem{Allen1999-Matrixformulationfo}
Allen L, Courtial J and Padgett M~J 1999 {\em Phys. Rev. E\/} {\bf 60}
  7497--7503 ISSN 2470-0053
  \urlprefix\url{https://doi.org/10.1103/PhysRevE.60.7497}

\bibitem{Albers2023-ASymplecticDynamics}
Albers P, Geiges H and Zehmisch K 2023 {\em Arnold Math. J.\/} {\bf 9} 41--68
  ISSN 2199-6806 \urlprefix\url{https://doi.org/10.1007/s40598-021-00195-7}

\bibitem{BibEntry2024Nov}
S R 2024 {\em cphysics.org\/}
  \urlprefix\url{https://www.cphysics.org/article/03838.pdf}

\bibitem{Bruzzo2023-D3-branesupergravity}
Bruzzo U, Fr{\ifmmode\acute{e}\else\'{e}\fi} P, Shahzad U and Trigiante M 2023
  {\em Lett. Math. Phys.\/} {\bf 113} 64--70 ISSN 1573-0530
  \urlprefix\url{https://doi.org/10.1007/s11005-023-01683-x}

\bibitem{Cisowski2026Feb}
Cisowski C 2026 {\em Phys. Rev. A\/} {\bf 113}(6) 063530
  \urlprefix\url{https://link.aps.org/doi/10.1103/gf21-tgcm}

\bibitem{Lee2022-TopologyandGeometry}
Lee C~H and Tan C~H 2022 {Topology and Geometry of 3-Band Models} {\em {IRC-SET
  2021}\/} (Singapore: Springer) pp 59--81 ISBN 978-981-16-9869-9
  \urlprefix\url{https://doi.org/10.1007/978-981-16-9869-9_5}

\bibitem{Schwinger2015-OnAngularMomentum}
Schwinger J 2015 {\em {On Angular Momentum (Dover Books on Physics)}\/}
  (Mineola, NY, USA: Dover Publications) ISBN 978-0-48678810-4
  \urlprefix\url{https://www.amazon.com/Angular-Momentum-Dover-Books-Physics/dp/0486788105}

\bibitem{Shabbir2017-MajoranaRepresentati}
Shabbir S 2017 {\em Majorana Representation in Quantum Optics : SU(2)
  Interferometry and Uncertainty Relations\/} Ph.D. thesis KTH Royal Institute
  of Technology
  \urlprefix\url{http://www.diva-portal.org/smash/record.jsf?pid=diva2%3A1091994&dswid=-7801}

\bibitem{Majorana1932-Atomiorientatiinca}
Majorana E 1932 {\em Nuovo Cim.\/} {\bf 9} 43--50 ISSN 1827-6121
  \urlprefix\url{https://doi.org/10.1007/BF02960953}

\bibitem{Gutierrez-Cuevas2020-ModalMajoranaSphere}
Guti{\ifmmode\acute{e}\else\'{e}\fi}rrez-Cuevas R, Wadood S~A, Vamivakas A~N
  and Alonso M~A 2020 {\em Phys. Rev. Lett.\/} {\bf 125} 123903 ISSN 1079-7114
  \urlprefix\url{https://doi.org/10.1103/PhysRevLett.125.123903}

\bibitem{Perumangatt2015Generalized}
Chithrabhanu P, Reddy S~G, Anwar A and Singh R~P 2015 {Generalized orbital
  angular momentum Poincar{\ifmmode\acute{e}\else\'{e}\fi} sphere} {\em
  {International Conference on Optics and Photonics 2015}\/} vol 9654 (SPIE) p
  409 \urlprefix\url{https://doi.org/10.1117/12.2182651}

\bibitem{Karimi2010PolarizationControlled}
Karimi E, Slussarenko S, Piccirillo B, Marrucci L and Santamato E 2010 {\em
  Phys. Rev. A\/} {\bf 81}(5) 053813
  \urlprefix\url{https://link.aps.org/doi/10.1103/PhysRevA.81.053813}

\bibitem{Yang2025May}
Yang J, Jiang J, Wang J and Zheng X 2025 {\em Adv. Phys. Res.\/} {\bf 4}
  2400083 ISSN 2751-1200 \urlprefix\url{https://doi.org/10.1002/apxr.202400083}

\bibitem{Palmerduca2025-Fournogotheorems}
Palmerduca E and Qin H 2025 {\em SciPost Phys.\/} {\bf 18} 188
  (\textit{Preprint} \eprint{2407.06276})
  \urlprefix\url{https://doi.org/10.21468/SciPostPhys.18.6.188}

\bibitem{Fernandez-Corbaton2012-Helicityandangular}
Fernandez-Corbaton I, Zambrana-Puyalto X and Molina-Terriza G 2012 {\em Phys.
  Rev. A\/} {\bf 86} 042103 ISSN 2469-9934
  \urlprefix\url{https://doi.org/10.1103/PhysRevA.86.042103}

\bibitem{Rao2026Jun}
Rao A~S 2026 {\em Adv. Photonics Res.\/} {\bf 7} e70220 ISSN 2699-9293
  \urlprefix\url{https://doi.org/10.1002/adpr.70220}

\bibitem{Lassen2009-ContinuousVariableEntanglement}
Lassen M, Leuchs G and Andersen U~L 2009 {\em Phys. Rev. Lett.\/} {\bf 102}(16)
  163602 \urlprefix\url{https://doi.org/10.1103/PhysRevLett.102.163602}

\bibitem{Holleczek2011May}
Holleczek A, Aiello A, Gabriel C, Marquardt C and Leuchs G 2011 {\em Opt.
  Express\/} {\bf 19} 9714--9736 ISSN 1094-4087
  \urlprefix\url{https://doi.org/10.1364/OE.19.009714}

\bibitem{Yi2015-Hybrid-orderPoincare}
Yi X, Liu Y, Ling X, Zhou X, Ke Y, Luo H, Wen S and Fan D 2015 {\em Phys. Rev.
  A\/} {\bf 91} 023801 ISSN 2469-9934
  \urlprefix\url{https://doi.org/10.1103/PhysRevA.91.023801}

\bibitem{Ling2016-Characterizationandmanipulation}
Ling X, Yi X, Dai Z, Wang Y and Chen L 2016 {\em J. Opt. Soc. Am. B\/} {\bf 33}
  2172--2176 \urlprefix\url{https://doi.org/10.1364/JOSAB.33.002172}

\bibitem{Ren2015-GeneralizedPoincare}
Ren Z~C, Kong L~J, Li S~M, Qian S~X, Li Y, Tu C and Wang H~T 2015 {\em Opt.
  Express\/} {\bf 23} 26586--26595 ISSN 1094-4087
  \urlprefix\url{https://doi.org/10.1364/OE.23.026586}

\bibitem{Tymchenko2015-GradientNonlinearPa}
Tymchenko M, Gomez-Diaz J~S, Lee J, Nookala N, Belkin M~A and
  Al{\ifmmode\grave{u}\else\`{u}\fi} A 2015 {\em Phys. Rev. Lett.\/} {\bf 115}
  207403 ISSN 1079-7114
  \urlprefix\url{https://doi.org/10.1103/PhysRevLett.115.207403}

\bibitem{suzuki2016commenthigherorderpancharatnamberry}
Suzuki M, Yamane K, Oka K, Toda Y and Morita R 2016 Comment on "higher order
  pancharatnam-berry phase and the angular momentum of light"
  (\textit{Preprint} \eprint{1602.02471})
  \urlprefix\url{https://arxiv.org/abs/1602.02471}

\bibitem{HatcherVBKT}
Hatcher A 2017 Vector bundles and k-theory available from the author's webpage

\bibitem{DavisKirkAlgebraicTopology}
Davis J~F and Kirk P 2001 {\em Lecture Notes in Algebraic Topology\/} (American
  Mathematical Society)

\bibitem{SANSONETTO2010501}
Sansonetto N and Spera M 2010 {\em Journal of Geometry and Physics\/} {\bf 60}
  501--512 ISSN 0393-0440
  \urlprefix\url{https://doi.org/10.1016/j.geomphys.2009.11.012}

\bibitem{PENNA199899}
Penna V and Spera M 1998 {\em Journal of Geometry and Physics\/} {\bf 27}
  99--112 ISSN 0393-0440
  \urlprefix\url{https://doi.org/10.1016/S0393-0440(97)00070-3}

\bibitem{Barbierigeometry}
Barbieri G, Frizyuk K and Spera M 2025 {\em Available at SSRN 5371800\/}

\bibitem{Bliokh_2019}
Bliokh K~Y, Alonso M~A and Dennis M~R 2019 {\em Reports on Progress in
  Physics\/} {\bf 82} 122401
  \urlprefix\url{https://doi.org/10.1088/1361-6633/ab4415}

\bibitem{Solyom2007}
S{\ifmmode\acute{o}\else\'{o}\fi}lyom J and
  Pir{\ifmmode\acute{o}\else\'{o}\fi}th A 2007 {\em {Fundamentals of the
  Physics of Solids}\/} (Berlin, Germany: Springer) ISBN 978-3-54072600-5

\bibitem{Tsimokha2022Apr}
Tsimokha M, Igoshin V, Nikitina A, Toftul I, Frizyuk K and Petrov M 2022 {\em
  Phys. Rev. B\/} {\bf 105} 165311
  \urlprefix\url{https://doi.org/10.1103/PhysRevB.105.165311}

\bibitem{Vektorfeld2026Jan}
Vektorfeld 2026 {Selection rules for scattering by nanostructures {$\vert$}
  Optics {$\vert$} Acoustics}
  \urlprefix\url{https://www.youtube.com/watch?v=YZ1U6y6Fq10}

\bibitem{Li2015Jun}
Li G, Chen S, Pholchai N, Reineke B, Wong P~W~H, Pun E~Y~B, Cheah K~W, Zentgraf
  T and Zhang S 2015 {\em Nat. Mater.\/} {\bf 14} 607--612 ISSN 1476-4660
  \urlprefix\url{https://doi.org/10.1038/nmat4267}

\bibitem{Tymchenko2015Nov}
Tymchenko M, Gomez-Diaz J~S, Lee J, Nookala N, Belkin M~A and
  Al{\ifmmode\grave{u}\else\`{u}\fi} A 2015 {\em Phys. Rev. Lett.\/} {\bf 115}
  207403 \urlprefix\url{https://doi.org/10.1103/PhysRevLett.115.207403}

\bibitem{[Perelomov86]}
Perelomov A 1986 {\em {Generalized Coherent States and Their Applications}\/}
  (Berlin, Germany: Springer) ISBN 978-3-642-61629-7
  \urlprefix\url{https://link.springer.com/book/10.1007/978-3-642-61629-7}

\bibitem{Kuratsuji2013Sep}
Kuratsuji H 2013 {\em Phys. Rev. A\/} {\bf 88} 033801
  \urlprefix\url{https://doi.org/10.1103/PhysRevA.88.033801}

\bibitem{Jisha2021-GeometricPhaseinOp}
Jisha C~P, Nolte S and Alberucci A 2021 {\em Laser Photonics Rev.\/} {\bf 15}
  2100003 ISSN 1863-8880 \urlprefix\url{https://doi.org/10.1002/lpor.202100003}

\bibitem{Born1928Mar}
Born M and Fock V 1928 {\em Z. Phys.\/} {\bf 51} 165--180
  \urlprefix\url{https://doi.org/10.1007/BF01343193}

\bibitem{Levay2005Sep}
L{\ifmmode\acute{e}\else\'{e}\fi}vay P 2005 {\em arXiv\/} (\textit{Preprint}
  \eprint{math-ph/0509064})
  \urlprefix\url{https://doi.org/10.48550/arXiv.math-ph/0509064}

\bibitem{DeZela2012Feb}
De~Zela F 2012 {The Pancharatnam-Berry Phase: Theoretical and Experimental
  Aspects} {\em {Theoretical Concepts of Quantum Mechanics}\/} (Croatia:
  IntechOpen) ISBN 978-953-51-0088-1
  \urlprefix\url{https://doi.org/10.5772/34882}

\bibitem{Weinberg}
Weinberg S {\em {The Quantum Theory of Fields, Volume 1: Foundations}\/}
  (Cambridge, England, UK: Cambridge University Press) ISBN 978-0-52167053-1
  \urlprefix\url{https://www.amazon.de/-/en/Quantum-Theory-Fields-Foundations/dp/0521670535}

\end{thebibliography}

\appendix

\section{Stereographic Projection}
    \label{app:stere}
\begin{figure}[ht]
    \centering
    \includegraphics[width=0.5\textwidth]{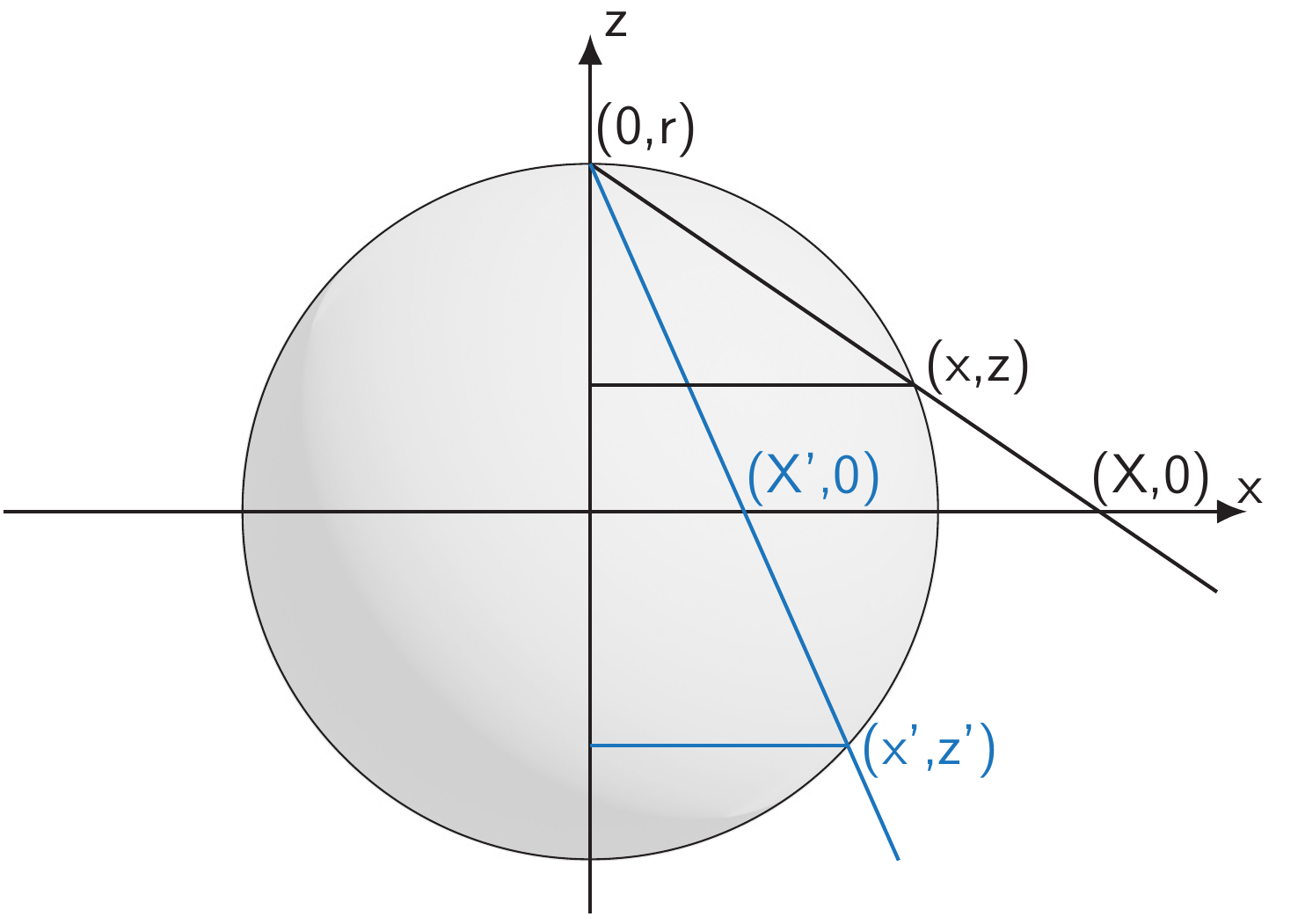}
    \caption{Stereographic projection establishes a one-to-one correspondence between the points on the sphere and those on the plane.}
\end{figure}
Let the center of a sphere of radius $r$ be at $(0,0,0)$.
Draw a line from the north pole that intersects the sphere and the plane $z=0$ at $(x,y,z)$ and $(X,Y,0)$, respectively.
We establish a relationship between the coordinates.
From the similarity of triangles,
\begin{equation}
    X=\frac{rx}{r-z}, \ \ \ 
    Y=\frac{ry}{r-z}.
\end{equation}
However, we would like to express the sphere coordinates in terms of the plane coordinates. For this, we consider the plane as a complex plane and introduce
\begin{equation}
    Z=\frac{r(x+\iu y)}{r-z}, \ \ \ 
    \overline Z=\frac{r(x-\iu y)}{r-z}.
\end{equation}
Consider
\begin{equation}
    Z\overline Z=\frac{r^2(x^2+y^2)}{(r-z)^2}=\frac{r^2(r^2-z^2)}{(r-z)^2}=\frac{r^2(r+z)}{(r-z)},
\end{equation}
from here
\begin{equation}
    z=\frac{r(Z\overline Z-r^2)}{(Z\overline Z+r^2)},
\end{equation}
\begin{equation}
    x=\frac{X(r-z)}{r}=\frac{(Z+\overline Z)(r-z)}{2r}=\frac{r^2(Z+\overline Z)}{(Z\overline Z+r^2)},
\end{equation}
\begin{equation}
    y=\frac{Y(r-z)}{r}=\frac{-\iu r^2(Z-\overline Z)}{(Z\overline Z+r^2)}
\end{equation}
Thus, every point on the sphere, except the north pole, corresponds to a point on the plane. The extended plane includes the point at infinity, which corresponds to the north pole.
Stereographic projection preserves angles and generalized circles; that is, it maps circles on the sphere to circles or straight lines on the plane.
\section{Projective space}
\label{app:proj}
One can also find a detailed explanation and relation to the Poincar\'e sphere in~\cite{vid2025Nov}.
The complex projective space $\mathbb{C}P^n$ is the quotient set $\mathbb{C}^{n+1} \backslash \{0\}$ with respect to the equivalence relation 
\begin{equation}
	x\sim y \Leftrightarrow \exists \lambda \in \mathbb{C} \backslash\{0\}\colon x=\lambda y
\end{equation}
\begin{equation}
	\mathbb{C}P^n=(\mathbb C^{n+1}\backslash \{0\})/{\sim}
\end{equation}
One can often encounter notations like $\mathbb C \backslash \{0\} \coloneq \mathbb C^*$. The complex projective line $\mathbb{CP}^1$ is homeomorphic to a sphere, which is also called the Riemann sphere. 
In other words, the Riemann sphere can be defined as the set of all lines passing through the origin in $\mathbb{C}^2$. 
Any point on the Riemann sphere can be represented by a pair of complex numbers $(z_1, z_2)$, and two points $(z_1, z_2)$ and $(w_1, w_2)$ represent the same point on the Riemann sphere if there exists a nonzero complex number $\lambda$ such that:
\begin{equation}
 	(z_1, z_2) = \lambda(w_1, w_2).   
\end{equation}
This equivalence relation introduces the concept of projective coordinates on the Riemann sphere, typically denoted as $[z_1 : z_2]$. 
The complex plane $\mathbb{C}$ corresponds to the subset of the Riemann sphere where $z_2 \neq 0$, and each point $z \in \mathbb{C}$ is identified with the point $[z_1/z_2 : 1]=[z : 1]$ on the Riemann sphere. The point at infinity is represented by $[1 : 0]$.

\section{Derivations for the SU(2) matrix for the rotated waveplate}
\label{app:waveplate}
The optical axis of the waveplate is rotated by an angle $\beta$ with respect to the $x$-axis.
We can introduce new $E_{x'}$ and $E_{y'}$ in another coordinate system along the symmetry planes of the waveplate.

In general case it is 
\begin{equation}
	E_{x} = \cos{\beta}E_{x'} - \sin{\beta}E_{y'}, \ \ \ \ \ 
	E_{y} =  \sin{\beta} E_{x'} +  \cos{\beta}E_{y'}
\end{equation}
The initial circularly polarized wave is expressed through the new coordinates as
\begin{equation}
	E_R = e^{-\iu \beta}\frac{E_{x'} - \iu E_{y'}}{\sqrt{2}}, \ \ \ \ 
	E_L = e^{\iu \beta}\frac{E_{x'} + \iu E_{y'}}{\sqrt{2}}
\end{equation}
\begin{equation}
	E_{x'} = \frac{e^{\iu \beta}E_{R} +  e^{-\iu \beta} E_{L}}{\sqrt{2}}, \ \ \ \ 
	E_{y'} = -\frac{e^{\iu \beta}E_{R} -  e^{-\iu \beta} E_{L}}{\iu \sqrt{2}}
\end{equation}
Therefore for electric field components transmitted through a rotated waveplate we have:
\begin{align}
	\tilde{E}_R &= e^{-\iu \beta}\left( e^{\iu\delta/2}\frac{e^{\iu \beta}E_{R} +  e^{-\iu \beta} E_{L}}2 + \iu  e^{-\iu\delta/2} \frac{e^{\iu \beta}E_{R} -  e^{-\iu \beta} E_{L}}{2\iu}\right) \\ 
	\tilde{E}_L &= e^{\iu \beta}\left( e^{\iu\delta/2}\frac{e^{\iu \beta}E_{R} +  e^{-\iu \beta} E_{L}}2 - \iu  e^{-\iu\delta/2} \frac{e^{\iu \beta}E_{R} -  e^{-\iu \beta} E_{L}}{2\iu}\right) \\ 
	\tilde{E}_R &=\left(e^{\iu\delta/2}\frac{E_{R} +   e^{-2 \iu \beta} E_{L}}2 + e^{-\iu\delta/2} \frac{E_{R} - e^{-2 \iu \beta}E_{L}}{2}\right) \\
	\tilde{E}_L &=\left(e^{\iu\delta/2}\frac{e^{2 \iu \beta}E_{R} +    E_{L}}2 -  e^{-\iu\delta/2} \frac{e^{2 \iu \beta}E_{R} - E_{L}}{2}\right)
\end{align}
\begin{equation} 
	\begin{pmatrix}
		\tilde{E}_R \\
		\tilde{E}_L 
	\end{pmatrix}=
	\begin{pmatrix}
		\cos \delta/2 & \iu e^{-2 \iu \beta}\sin \delta/2 \\
		\iu  e^{2 \iu \beta}\sin \delta/2 & \cos \delta/2
	\end{pmatrix}
	\begin{pmatrix}
		E_R \\
		E_L 
	\end{pmatrix} 
\end{equation}
So we get the general formula:
\begin{equation} 
	\begin{pmatrix}
		\tilde{E}_R \\
		\tilde{E}_L 
	\end{pmatrix}=
	(\mathbb I \cos \delta/2 + \iu (\cos 2\beta\sigma_x+\sin 2\beta\sigma_y) \sin \delta/2 )
	\begin{pmatrix}
		E_R \\
		E_L 
	\end{pmatrix}.
    \label{446matrix}
\end{equation}

  \subsection{Horizontal lift and holonomy: an example}
\label{app:lift}
For more general considerations see also the book~\cite{Nakahara}.
Let $\mathcal A$ be the global $U(1)$ connection $1$--form on $S^3$.
For a lifted curve $z(t)\in S^3\subset\mathbb C^2$ the horizontality is
\begin{equation}
\label{eq:horiz}
\mathcal A \big(\dot z(t)\big)=0.
\end{equation}
Here we use $z=(z_1,z_2)$ with $z_1=x_1 + \iu x_2$, $z_2=x_3 + \iu x_4$.
Let us fix any (not necessarily horizontal, preferably closed) reference lifted curve $z_{\rm ref}(t)$
with the same base projection.
Any other lift differs by a fibre rotation $e^{-\iu\alpha(t)}$, hence
\begin{equation}
\label{eq:decomp}
    z(t) = e^{-\iu\alpha(t)} z_{\rm ref}(t),\qquad
    \dot z = e^{-\iu\alpha(t)}\big(\dot z_{\rm ref} -  \iu\dot\alpha  z_{\rm ref}\big).
\end{equation}
Using the $U(1)$--invariance of $\mathcal A$ and \blue{$\mathcal A(u_v)= \iu$,}
the horizontality condition \eqref{eq:horiz} with \eqref{eq:decomp} gives
\begin{align}
\label{eq:alpha-ode}
    0 = \mathcal A(\dot z) = \mathcal A(\dot z_{\rm ref}) -\blue{ \iu \dot\alpha(t)}
    \quad\Longrightarrow\\ \nonumber
    {\dot\alpha(t) = \blue{-\iu \mathcal A \big(\dot z_{\rm ref}(t)\big)}}.
\end{align}
In real coordinates on $S^3\subset\mathbb R^4$ we use
\begin{align}
    \label{eq:A-R4}
    \mathcal A = \blue{\iu(x_1 \dd x_2 - x_2\dd x_1 + x_3\dd x_4 - x_4\dd x_3)},
   \\ \nonumber
    \mathcal A(\dot x)= \blue{\iu(x_1\dot x_2-x_2\dot x_1+x_3\dot x_4-x_4\dot x_3)}.
\end{align}

Now we consider the example of the closed curve on the Poincar\'e sphere with a fixed $\theta$ and denote
$a=\cos(\theta/2)$, $b=\sin(\theta/2)$.
A convenient (non-horizontal) closed reference lift of this parallel is given by
\begin{equation}
    \label{eq:xref-hopf}
    x_{\mathrm{ref}}(t)
    = \big(a,\; 0,\; b\cos\varphi(t),\; -b\sin\varphi(t)\big),
\end{equation}
whose Hopf projection is the latitude (parallel) on $S^2$ with $Z=\cos\theta$.
Indeed,
$ z_1(t)=a$, 
$ 
z_2(t)=b\cos\varphi(t)-\iu b\sin\varphi(t)=b\,e^{-\iu\varphi(t)},
$
and using the Hopf map~\eqref{eq:projection}
we obtain $z_1\overline{z_2}=a\cdot b\,e^{\iu\varphi(t)},$
\[
\pi\big(z_{\rm ref}(t)\big)=\big(\sin\theta\cos\varphi(t),\,\sin\theta\sin\varphi(t),\,\cos\theta\big).
\]
Along this curve the connection form evaluates \blue{to
\begin{align}
\label{eq:A-on-xref-hopf}
\mathcal{A} \big(\dot{x}_{\mathrm{ref}}(t)\big)
= -\iu b^2\,\dot{\varphi}(t) = \\ \nonumber
= -\iu\sin^2\!\frac{\theta}{2}\,\dot{\varphi}(t)
= -\frac{\iu}{2}\big(1-\cos\theta\big)\dot{\varphi}(t).
\end{align}}
Therefore, the horizontality equation \eqref{eq:alpha-ode} gives
\begin{equation}
\dot\alpha(t)=\blue{-\iu \mathcal{A}}\big(\dot{x}_{\mathrm{ref}}(t)\big)
= -\frac12\big(1-\cos\theta\big)\dot{\varphi}(t).
\end{equation}
For a closed loop along the parallel, $\varphi$ increases by $2\pi$, hence
\begin{align}
\label{eq:holonomy-hopf}
\Delta\alpha
= - \frac12 \int_0^{2\pi}\big(1-\cos\theta\big)\dd\varphi
= - \pi\big(1-\cos\theta\big),\\ \nonumber
\mathrm{Hol}=\exp\big(-\iu\Delta\alpha\big).
\end{align}
Since the solid angle of the spherical cap bounded by the parallel is
\(\displaystyle \Omega=2\pi(1-\cos\theta)\), we have the equivalent relation
\begin{equation}
\label{eq:holonomy-solidangle}
\Delta\alpha \equiv -\frac12 \Omega,
\qquad
\mathrm{Hol}=\exp\left(\iu\frac{\Omega}{2}\right).
\end{equation}
Note that we could consider a different reference lift, \textit{e.g.}
\begin{align}
    \label{eq:xref-hopf2}
    &x_{\mathrm{ref}}(t)
    = \\ =\nonumber &\left(a\cos\frac{\varphi(t)}{2}, a\sin\frac{\varphi(t)}{2}, b\cos\frac{\varphi(t)}{2}, -b\sin\frac{\varphi(t)}{2}\right),
\end{align}
and this one is not closed. 
This means that one should add an additional phase, in this case, $\pi$, which corresponds to the non-closedness of the reference lift. 
This is related to the part of the phase that is referred to as dynamical in~\cite{Bliokh_2019}.
\subsection{Global and local descriptions}
In practical applications one usually does not work directly with the global connection form $\mathcal{A}$ on the total space $E$, but rather with its local representatives on the base manifold. 
This step is often left implicit, but it is important to state it explicitly: obtaining a local connection on the base requires the choice of a local section of the fibre bundle.
We introduce angular coordinates on $S^3$ $(\theta,\varphi,\psi)$ with
$\theta\in[0,\pi]$,  $\varphi\in[0,2\pi)$, and $\psi\in[0,4\pi)$,
and write~\cite{Chruscinski-GeometricPhasesinC}
\begin{align}
    z_1 &= \cos\frac{\theta}{2}\,\exp\left(\iu \frac{\psi+\varphi}{2}\right), \label{eq:Hopf_z1}\\
    z_2 &= \sin\frac{\theta}{2}\,\exp\left(\iu \frac{\psi-\varphi}{2}\right). \label{eq:Hopf_z2}
\end{align}
The projection $\pi\colon S^{3}\to S^{2}$ given by the Hopf map gives
\begin{equation}
    x = \sin\theta\cos\varphi,\quad
    y = \sin\theta\sin\varphi,\quad
    z = \cos\theta ,
\label{eq:Hopf_projection}
\end{equation}
which is the standard spherical parametrization of $S^{2}$. The angles $(\theta,\varphi)$ therefore parametrize the base $B=S^{2}$, while $\psi$ parametrizes the fibre $F=S^{1}$.

To pass from the global connection on $E$ to a local gauge potential on $B$, we need a {local section}. 
A (local) section is a smooth map
\begin{equation}
    s \colon U \subset B \to E
\end{equation}
such that $\pi\circ s = \mathrm{id}_U$. 
For the Hopf bundle we use the canonical sections
\begin{align}
    f_N(\theta,\varphi) &= (\theta,\varphi,\psi=-\varphi), \label{eq:Hopf_sec_N}\\
    f_S(\theta,\varphi) &= (\theta,\varphi,\psi=+\varphi), \label{eq:Hopf_sec_S}
\end{align}
defined on the coordinate patches
$U_N = S^2 \setminus \{\text{south pole}\},$  
$U_S = S^2 \setminus \{\text{north pole}\},$
respectively, so that each section is smooth on its domain.

Given a connection 1-form $\mathcal{A}$ on $E$, its local representative on $U\subset B$ is obtained as a pullback along a chosen section $s$. 
Let $s\colon B\to E$ be a smooth map and let $\alpha$ be a 1-form on $E$. 
The \textbf{pullback} of the 1-form $\alpha$, denoted by $s^{*}\alpha$, is the 1-form on $B$ defined by
\begin{equation}
    (s^{*}\alpha)_x(v) = \alpha_{s(x)}(s_{*}v),
    \label{eq:pullback_1form}
\end{equation}
for any $x\in B$ and $v\in T_xB$, where $s_{*}\colon T_xB\to T_{s(x)}E$ is the pushforward.
In spherical coordinates the global connection~\eqref{eq:A_in4cord} is written as
\begin{equation}
    \mathcal{A} = \iu g \bigl(\dd\psi + \cos\theta\, \dd\varphi\bigr).
    \label{eq:Hopf_global_connection}
\end{equation}
The local connection forms (also called gauge potentials) on $U_N$ and $U_S$ are then defined by
\[
A_N := f_N^{*}\mathcal{A}, \qquad A_S := f_S^{*}\mathcal{A}.
\]
Using \eqref{eq:Hopf_sec_N}–\eqref{eq:Hopf_global_connection} and the fact that $f_N$ and $f_S$ only change $\psi$, we obtain
\begin{align}
A_N = f_N^{*}\mathcal{A}
    = \iu g\bigl(\dd(-\varphi) + \cos\theta\, \dd\varphi\bigr) \nonumber =  \\
    = - \iu g (1 - \cos\theta)\, \dd\varphi, \label{eq:AN}\\
A_S = f_S^{*}\mathcal{A}
    = \iu g\bigl(\dd\varphi + \cos\theta\, \dd\varphi\bigr) \nonumber = \\ 
    = \iu g (1 + \cos\theta)\, \dd\varphi. \label{eq:AS}
\end{align}
These 1-forms $A_N$ and $A_S$ are the local connection forms (gauge potentials) on the base $S^{2}$ in the northern and southern charts.
On the overlap $U_N\cap U_S$ they are related by the gauge transformation.

\subsection{Curvature and Chern classes}
\label{app:curva}
A crucial notion is then that of {\bf curvature} of a connection. 
Given two \textit{horizontal} vector fields $u$, $v$, their Lie bracket $([u,v](f) := u\bigl(v(f)\bigr) - v\bigl(u(f)\bigr))$ is, in general, {\it not horizontal}: the operator
\begin{equation}
    \blue{{\mathcal F} (u, v) := -{\mathcal A} ([u, v])}
\end{equation}
is the {\it curvature} of the given connection.
Geometrically, $\mathcal F(u,v)$ represents the infinitesimal holonomy associated with a small loop spanned by $u$ and $v$.
\par
The curvature of a connection is defined globally as the 
2-form, 
\begin{equation}
    \mathcal F := \dd \mathcal A + \mathcal A \wedge \mathcal A ,
\end{equation}
where
$
\dd \mathcal A(u,v)
=
u\bigl(\mathcal A(v)\bigr) - v\bigl(\mathcal A(u)\bigr) - \mathcal A\bigl([u,v]\bigr)
$
for \textit{any} vector fields $u,v$ on $E$ and $(\mathcal A \wedge \mathcal A)(u,v) = [\mathcal A(u),\mathcal A(v)]$. 
When dealing with the case of a {\bf principal bundle} --- wherein $F = G$, a Lie group, called {\it structure group} freely acting on fibres, namely, without fixed points --- and if $G$ is abelian,
 the second term will actually {\it drop out}. So in our case $G = U(1)$ and $\mathcal F = \dd \mathcal A$.
Stokes' theorem then yields (pulling back the relevant forms to the base via a local section $s$ so that, slightly abusively, we may look at $\mathcal{F}$ as a 2-form $F := s^*{\mathcal F}$ thereon):
\begin{equation}
    \int_{\mathscr C} {A} = \int_{\mathscr D} \dd {A} = \int_{\mathscr D} F  
\end{equation}
for a loop ${\mathscr C}$ bounding a 2-dimensional domain ${\mathscr D}$ on the base space and where $A = s^*{\mathcal A}$.
The holonomy $\mathrm{Hol}_{\mathcal{A}}({\mathscr C})$ of the connection $\mathcal{A}$ around ${\mathscr C}$, \textit{i.e.} the (generalized) Berry phase will then be
\begin{equation}
\mathrm{Hol}_{\mathcal{A}}({\mathscr C}) = \exp \left(-\int_{\mathscr C} {A}\right) = \exp \left(-\int_{\mathscr D} {F}\right) 
\end{equation}

Computing  $F = \dd A$ from either local potential \eqref{eq:AN} and \eqref{eq:AS} gives
\begin{align}
F_N = \dd A_N
    = - \iu g\, \dd(1 - \cos\theta)\wedge \dd\varphi \nonumber = \\
    = - \iu g\, \sin\theta\, \dd\theta \wedge \dd\varphi, \label{eq:FN}\\
F_S = \dd A_S
    = \iu g\, \dd(1 + \cos\theta)\wedge \dd\varphi \nonumber = \\
    = - \iu g\, \sin\theta\, \dd\theta \wedge \dd\varphi. \label{eq:FS}
\end{align}
Thus $F_N = F_S =: F$ defines a globally well-defined curvature 2-form on $S^2$.
For a closed curve ${\mathcal C}\subset S^2$ bounding an oriented surface
${\mathcal D}$, Stokes' theorem yields
\begin{equation}
    \oint_{\mathcal C} A_{N/S} = \iint_{\mathcal D} F.
\end{equation}
Since $\sin\theta\, \dd\theta \wedge \dd\varphi$ is the area form on the unit
sphere, the integral of the curvature over ${\mathcal D}$ equals the solid angle
$\Omega({\mathcal C})$ enclosed by ${\mathcal C}$:
\begin{equation}
    \iint_{\mathcal D} F = - \iu g\, \Omega({\mathcal C}).
\end{equation}
Finally, integrating the curvature over the whole sphere gives
$\iint_{\mathcal D} F = -4\pi \iu g$.
The first Chern number therefore reads
\begin{equation}
c_1
=
\frac{1}{-2\pi \iu}\iint_{S^2} F
=
2g,
\end{equation}
which reduces to $c_1=1$ upon choosing the standard normalization $g=\tfrac12$.
\section{SU(2)-actions and parallel transport}

\label{app:rightleft}
We use again $z_1=x_1+\iu x_2$ and $z_2=x_3+\iu x_4$.
A vector field is written in the real basis $(\partial_{1},\ldots,\partial_{4})$.
Let $U\in SU(2)$ act on every point of $\mathbb C^2$ $(z_1, z_2)$ as follows:
\begin{equation}
    \Phi_U:\ (z_1,z_2)^\mathsf{T} \mapsto (z_1',z_2')^\mathsf{T}=U\,(z_1,z_2)^\mathsf{T}.
\end{equation}
This induces a real linear map $Q\in SO(4)$ (compare with rotation in quaternion form) on $x=(x_1, x_2, x_3,x_4)^\mathsf{T}$:
\begin{equation}
    x' = Q\,x.
\end{equation}
We want to find a vector field on $S^3$ that corresponds to an infinitesimal rotation generated by such a unitary matrix.
Now consider the example $U=\cos \delta/2 \ \mathbb I+\iu\sin \delta/2 \  \sigma_x$. For this case
\begin{equation}
    Q_x = 
    \begin{pmatrix}
        \cos \delta/2 &0&0&-\sin \delta/2\\
        0&\cos \delta/2&\sin \delta/2&0\\
        0&-\sin \delta/2&\cos \delta/2&0\\
        \sin \delta/2&0&0&\cos \delta/2
    \end{pmatrix}.
\end{equation}
In the general case~\eqref{su2matrix} we introduce the notation
\begin{align}
    c=\cos \delta/2,\ s=\sin \delta/2,\\
    c_{2}=\cos(2\beta),\ s_{2}=\sin(2\beta)
\end{align}
and get 
\begin{equation}
    Q(\delta,\beta)=
    \begin{pmatrix}
        c & 0 & s\,s_{2} & -s\,c_{2} \\
        0 & c & s\,c_{2} & s\,s_{2} \\
        -s\,s_{2} & -s c_{2} & c & 0 \\
        s\,c_{2} & -s\,s_{2} & 0 & c
    \end{pmatrix},
\end{equation}
{and with an additional $\sigma_z$ rotation $U=\cos \delta/2 \ \mathbb I+\iu\sin \delta/2 \  \sigma_z$, which is not realized via the waveplate
\begin{equation}
    Q_z=
    \begin{pmatrix}
        c & -s & 0 & 0 \\
        s & c & 0 & 0 \\
        0 & 0 & c & s \\
        0 & 0 & -s & c
    \end{pmatrix}.
\end{equation}
When considering an infinitesimal rotation we should just take the corresponding generator $L_Q$ to get the corresponding vector field as $\dot x = L_Q x$, leading to, \textit{e.g.} 
\begin{equation}
    L_z=
    \begin{pmatrix}
        0 & -1 & 0 & 0 \\
        1 & 0 & 0 & 0 \\
        0 & 0 & 0 & 1 \\
        0 & 0 & -1 & 0
    \end{pmatrix}.
\end{equation}
with corresponding vector field 
\begin{equation}
    u_z = -x_2\partial_1 + x_1\partial_2 + \{x_4\partial_3 - x_3\partial_4\}.
\end{equation}
}
Analogously, 
\begin{equation}
    u_x = -x_4\partial_1 {+} \{x_3\partial_2 - x_2\partial_3\} + x_1\partial_4,
\end{equation}
\begin{equation}
    u_y = \{x_3\partial_1\} + x_4\partial_2 - \{x_1\partial_3\} - x_2\partial_4.
\end{equation}
Note that they differ from  horizontal or vertical vector fields~\eqref{eq:uv},~\eqref{eq:uh12}. 
{For convenience, we put the non-matching terms in curly brackets.
}
So, a unitary 1-parameter subgroup
realizes a specific motion on $S^3$ along the vector fields, given above, which are not horizontal. 
This means that, in general, a point in $S^3$ does not move in the way prescribed by the connection form~\eqref{eq:connec}, and when we have a closed curve on the Poincar\'e sphere, realized by a $SU(2)$ rotation, the lifted curve on the $S^3$ sphere is not horizontal (see, for example, Fig. 3 in~\cite{Courtial1999Dec} as an instance of non-parallel transport). 
However, for particular lines, where non-coinciding terms for $u_x$ and $u_y$ are zero, we still move along  horizontal vector fields! 
For example, for $u_x$ we want $x_3=x_2=0$, which means that $z_1$ is purely real and $z_2$ is purely imaginary. 
That corresponds to the great circle with $x=0$~\eqref{eq:projection}.
But $\sigma_x$ corresponds to the rotation of the Poincar\'e sphere around the $x$-axis, which means that the motion in $S^3$
induced by that along a great circle in $S^2$ is  indeed dictated by parallel transport.  
Similarly, we can consider other rotation angles of the waveplate, however, moving on a great circle of $S^2$ provides a necessary condition for the phase to be geometric. 
\section{Eigenmodes, scattering and irreps}
\label{app:modes}
In this section we discuss, how the shape of the nanostructures was chosen based on their scattering properties for a particular vortex beam.
{As follows from Wigner's theorem~\cite{Solyom2007}, if a resonator belongs to a particular symmetry group, \textit{i.e.}, it does not change its shape under all transformations of that group, its eigenmodes transform according to its irreducible representations. 
This means that eigenmodes of different symmetries can be classified according to the irreducible representations (irreps)~\cite{Gladyshev_Frizyuk_Bogdanov_2020, Xiong:20}.
The symmetry behavior of the scattered wave can also be decomposed into a direct sum of irreps, and a particular mode will be excited only if the same irrep belongs to the decomposition of the incident wave~\cite{Gladyshev_Frizyuk_Bogdanov_2020, Tsimokha2022Apr}. 

However, in our case of axially symmetric systems, these considerations may be simplified. 
We are interested only in a single quantity that describes the symmetry behavior of the incident wave and the eigenmodes under rotations around the z-axis: the projection of the total angular momentum. 
Plane waves and the discussed vortex beams possess a well-defined value of $m$. 
But then the nanostructure has $\mathfrak{n}$-fold symmetry. 
This means that each eigenmode contains all possible values of $m$ that differ by $\mathfrak{n}\nu, \  \nu \in \mathbb Z$. 
This means that the function describing this eigenmode may be decomposed into a series, each term of which has the same symmetry behavior as the exponential $\text{e}^{\iu (m_i+\mathfrak{n}\nu) \varphi}$, where $m_i$ is an integer that characterizes the eigenmode.
It can be understood with the help of a Fourier-like decomposition of the shape of the nanostructure, written as a function of the angle $\varphi$. 
Rigorous considerations for vector functions can be found in~\cite{Gladyshev_Frizyuk_Bogdanov_2020}, and a simplified scalar version in~\cite{Tsimokha2022Apr, Vektorfeld2026Jan}. Thus, after all, if $m$ is the TAM projection of the incident field, then all eigenmodes for which
\begin{equation}
    \exists \nu \in \mathbb Z,\  m=m_i+\mathfrak{n}\nu
\end{equation}
is valid are excited.

\blue{In particular,} this means that if we have a triangular nanostructure, $\mathfrak{n}=3$, and, for example, the incident beam has $m=3$, then the $m=-3$ TAM projection \blue{ could} also contribute to the scattered field, because for the triangular structure, all eigenmodes with TAM projection $3+3\nu$ are excited, and for the hexagonal structure, $3+6\nu$. However, not only does $m=-3$ contribute to scattering, but also an infinite number of possible values.
 
The contribution of each TAM projection depends on the resonances and the particular shape of the structure and can be optimized.

Another way to view this is to keep in mind that two different eigenmodes, or more precisely two different types of eigenmodes, must be excited with different phases. 
They should be independently excited by
\begin{align}
	   \ket{B_e} = \frac{\ket{R} + \ket{L}}{\sqrt{2}} \text{ and }
      \ket{B_o} =  -\iu\frac{\ket{R} - \ket{L}}{\sqrt{2}},
\end{align}
with different phases and amplitudes. 
This would be analogous to horizontal and vertical linear polarizations on the Poincaré sphere. This means that such eigenmodes should each transform according to a 1-dimensional irrep, because if they transformed according to a 2-dimensional irrep into each other, they would have the same amplitudes and phases. Group representation theory tells us that this is valid for shapes for which the condition $2m=\mathfrak{n}\nu$ is satisfied, exactly as in the previous case.
}
\subsection{Cases with different TAM projections and another view on the geometry}
{The higher-order Poincar\'e sphere is defined based on two basis states, and in principle, it does not necessarily have to be based on mirror-symmetric states, such as left- and right-polarized plane waves or the previously considered beams. 
We can construct something similar for any two arbitrary values of $m$. 
Considerations based on irreps are not straightforward in this case and can be misleading, because now the modes can transform according to any irrep. 
However, the selection rules, which determine whether a mode with another $m$ will be excited, are still valid: 
\begin{equation}
  m^\text{inc}-m^\text{out}=\mathfrak n\nu.  
\end{equation}
But how can we now determine, which phase is obtained during rotation of the nanostructure?
We can, in principle, follow the same approach as described in~\cite{guercio2026tensordrivengeometricphasenonlinear, Li2015Jun, Tymchenko2015Nov}. 
For this, we consider the incident-field--nanostructure--particular-scattering-TAM-component system and first rotate the system as a whole by some angle $\beta$. 
The incident and scattered fields acquire the phases $e^{-\iu m^{\text{out/inc}} \beta}$, each with its own value of $m$. 
Then, we perform a time shift such that the incident field coincides with the one before rotation. 
This corresponds to the same incident field, but a rotated nanostructure. 
The scattered field already has the phase $e^{-\iu m^\text{out} \beta}$, and after the time shift it acquires an additional phase $e^{\iu m^\text{inc} \beta}$, because it has the same frequency as the incident field. This gives the expression~\eqref{eq:phasem}. Note that here we did not use the language of unitary matrices at all, nor did we consider any real path on the Poincar\'e sphere and its holonomy.

The geometric interpretation of this phase remains quite formal and may be somewhat forced. It again raises the question of the definition of the geometric phase in physics: should the holonomy interpretation be necessary, or can one just rely on the vague idea that ``the phase depends somehow only on the geometry, but not on the resonances'', which, of course, does not yet constitute a rigorous definition but may possibly be refined.
}

{
\section{Interpretation in terms of coherent states}
The preceding observations could be rephrased via Perelomov's coherent states (\cite{[Perelomov86],Gutierrez-Cuevas2020-ModalMajoranaSphere, Kuratsuji2013Sep}, see also \cite{Barbierigeometry}. In physical terms, let us choose a state on the Poincaré sphere, for instance, a pole.
\blue{Any transformation that belongs} to the group $SU(2)$ can either move the state to another one, which corresponds to a different point on the sphere, or just introduce some phase (for the pole it would be the subgroup of rotations around the $z$-axis).
In mathematical terms, we consider an action of a compact Lie group $G$ ($SU(2)$ in our case) on a finite-dimensional vector space $V$ (space of polarization states)
and obtain the orbit of this action on a vector $|\xi\rangle \in V$:
\begin{equation}
    \mathcal O_\xi =
    \{ [U(g)|\xi\rangle] \,|\, g\in G \}
    \cong G/H .
\end{equation}
where $H$, called the isotropy group of $ |\xi\rangle$ (in our case $H$ is $U(1)$ and $G/H$ is topologically $S^2$),
satisfies
\begin{equation}
    U(h) |\xi\rangle = e^{\iu \{  \alpha (h)\}} |\xi\rangle
\end{equation}
with $e^{ \iu \alpha(\cdot)}$ a character of $H$, namely, a one-dimensional representation of $H$:
\begin{equation}
    e^{\iu \{  \alpha (h_1h_2)\}} = e^{\iu \{  \alpha (h_1)\}} \, e^{\iu \{  \alpha (h_2)\}}, \quad h_1, h_2 \in H.
\end{equation} 
The above arrangement is termed a {\it coherent state system associated with the representation $U(\cdot)$ and the vector
$|\xi\rangle$.} 
Physically, coherent states are all possible states which can be obtained from a given one by an action of the group.
Ultimately, we have a principal bundle with structure group $H$ and base manifold $G/H$.
\par
A unitary irreducible representation $U(\cdot)$ of $G = SU(2) \cong S^3$ is labelled by its ``spin'' $J$ and it has dimension $2J+1$. 
\blue{For the Poincar\'e-sphere orbit considered here,
$H=U(1)\cong SO(2)\approx S^1$.
Upon taking the corresponding reference vector
$|\xi\rangle\in V_J$ and acting thereon via the representation,
we obtain the coherent-state manifold}
\begin{equation}
    \{ [U(g)|\xi\rangle] \,|\, g\in SU(2) \}
    \cong SU(2)/U(1) \cong S^2 .
\end{equation}
}
{
\section{Details of numerical simulations}

\label{sec:num_det}

We numerically calculate the phase $\phi$ of the electric field acquired by rotation of a silicon particle exhibiting different symmetries under illumination by a circularly polarized paraxial Laguerre-Gaussian beam
\begin{equation}
    \vec{E}_{\text{inc}} \propto (\vec{\hat x}\pm\iu \vec{\hat y})u_{lp}^{\text{LG}}(\vec{r}).
\end{equation}
TAM projection $m$ of such an input field can be obtained by writing the expression for a circularly polarized field in equivalent form (see Eqs.~\eqref{eq:cyl_dec}, \eqref{eq:lg_expr}, \eqref{eq:lg_expr2})
\begin{equation}
    \vec{E}_{lp}^{\pm} = e^{\pm \iu \varphi} (\hat{\bm \rho}\pm \iu \hat{{\bm \varphi}}) u_{lp}^{LG}(\vec{r}).
\end{equation}
Combining the angular dependencies $e^{\pm\iu\varphi}e^{\iu l\varphi}$, for the TAM projection of an LG beam we have $m = l\pm1$. The phase acquired by the rotation of the particle under excitation with a beam ($\lambda = 1550$ nm) having TAM-projection $m$ was determined as follows:
\begin{equation}
    \phi_{\pm m} = \arg\left[\int_S \vec{E}(\rho,\varphi,z_0)\cdot (\hat{\bm \rho}\mp \iu \hat{{\bm \varphi}})e^{\mp \iu m\varphi} \text{d}S \right],
\end{equation}
where the upper(lower) sign corresponds to the right(left) hand light helicity; the integration was performed over the surface positioned below the particle ($z_0 = -1$~$\mu$m). 
Geometrical parameters of the particles were the following: $D_{2h}$ particle width $w=263$ nm, height $h = 790$ nm, depth $d = 345$ nm; $D_{6h}$ particle $w=291$ nm, $h = 700$ nm, the radius of the inscribed circle $R= 922$ nm; $D_{3h}$ particle $w=591$ nm, $h = 700$ nm, $R= 1300$ nm.
Calculations were performed using COMSOL Multiphysics\texttrademark \  software.

}

\section{Horizontal lift and the usual Berry integral}
\label{sec:horizontal-berry-integral}
{In this section, we provide a correspondence between the standard notation, \textit{e.g.}, in~\cite{Berry1987-TheAdiabaticPhasea, Jisha2021-GeometricPhaseinOp, Born1928Mar}, and the present approach.}
{
Let $z=(z_1,z_2)\in S^3\subset\mathbb C^2$ be a normalized representative of a polarization state. We use the Hermitian product~\eqref{innerproduct}:
$
    \langle z|w\rangle=\overline{z_1}w_1+\overline{z_2}w_2.
$
Let $z\in S^3\subset\mathbb C^2$ be a normalized representative of a \textit{ray} (ray  $[z]\in\mathbb CP^1$ is the physical state defined up to an overall phase~\cite{Levay2005Sep, DeZela2012Feb, Weinberg}),
\begin{equation}
    \langle z|z\rangle=1.
\end{equation}
Consider another normalized representative $z'$ of a neighbouring ray. 
The Pancharatnam condition says that the two states are in phase if~\cite{Pancharatnam1956-Generalizedtheoryof, DeZela2012Feb}
\begin{equation}
    \langle z|z'\rangle\in\mathbb R_+.
\end{equation}
Now let the neighbouring representative be obtained from a smooth curve $z(t)\in S^3$,
\begin{equation}
    z'=z(t+\dd t)=z(t)+\dot z(t)\dd t.
\end{equation}
Then
\begin{equation}
    \langle z(t)|z(t+\dd t)\rangle =
    1+\langle z(t)|\dot z(t)\rangle\dd t.
\end{equation}
Since the curve stays on $S^3$, we have
\begin{equation}
    \frac{\dd}{\dd t}\langle z(t)|z(t)\rangle=0, \text{  therefore, }
    \langle\dot z(t)|z(t)\rangle+\langle z(t)|\dot z(t)\rangle=0,
\end{equation}
and hence
\begin{equation}
    \operatorname{Re}\langle z(t)|\dot z(t)\rangle=0.
\end{equation}
Thus $\langle z(t)|\dot z(t)\rangle$ is purely imaginary. 
The Pancharatnam condition requires
\begin{equation}
\langle z(t)|z(t+\dd t)\rangle\in\mathbb R_+.
\end{equation}
Using the previous expressions, we get both
$
    \operatorname{Im}\langle z(t)|\dot z(t)\rangle=0.
$
and
$
    \operatorname{Re}\langle z(t)|\dot z(t)\rangle=0,
$
which means
\begin{equation}
    \langle z(t)|\dot z(t)\rangle=0.
\end{equation}
This is precisely the condition that the tangent vector $\dot z(t)$ belongs to the horizontal space defined in Eq.~\eqref{eq:horcond}, and can be also found in, \textit{e.g.},~\cite{Simon1983-HolonomytheQuantum}.
Therefore, the horizontal space consists of those infinitesimal changes of the representative which keep neighbouring states in phase in the Pancharatnam sense.

Note, that (compare with~\eqref{eq:horiz})
\begin{equation}
    \mathcal A(\dot z) =
   \blue{\iu( x_1\dot x_2-x_2\dot x_1+x_3\dot x_4-x_4\dot x_3) =
   \iu} \operatorname{Im}\langle z|\dot z\rangle = \langle z|\dot z\rangle.
\end{equation}
The same object is usually written in Berry's notation after choosing a local
reference lift. 
Let the base point be denoted by $\mathbf R$, and
let $z_{\rm ref}(\mathbf R)$ be a chosen normalized representative of the
corresponding ray. 
For a curve \(\mathbf R(t)\) on the base, the reference lift is
\begin{equation}
    z_{\rm ref}(t)=z_{\rm ref}(\mathbf R(t)),
    \qquad
    \left\langle z_{\rm ref}(t)\middle|\dot z_{\rm ref}(t)\right\rangle =
    \left\langle z_{\rm ref}(\mathbf R)\middle|
    \nabla_{\mathbf R}z_{\rm ref}(\mathbf R)
    \right\rangle\cdot\dot{\mathbf R}.
\end{equation}
Therefore, in this local gauge, the usual Berry phase along a closed curve
$C$ is
\begin{equation}
    \gamma_{\rm B}[C] =
    \iu\oint_C
    \left\langle z_{\rm ref}(\mathbf R)\middle|
    \nabla_{\mathbf R}z_{\rm ref}(\mathbf R)
    \right\rangle\cdot \dd \mathbf R .
\end{equation}
Equivalently,
\begin{equation}
    \gamma =
    \iu\oint_C\langle z_{\rm ref}|\dd z_{\rm ref}\rangle.
\end{equation}
This is the same connection as above. }

\end{document}